\documentclass[a4paper,12pt]{article}

\usepackage{graphicx}
\usepackage{amssymb,amsmath}
\usepackage{amsfonts}

\numberwithin{equation}{section}

\begin{document}
\title{Self-consistent solution of Kohn-Sham equations for infinitely extended systems with inhomogeneous electron gas}
\author{D. V. Posvyanskii and A. Ya. Shul'man \\ Kotel'nikov Institute of Radio Engineering and Electronics of RAS, \\ Moscow, 125009 Russia}
\date{}
\maketitle

\begin{abstract}
The density functional approach in the Kohn-Sham approximation is
widely used to study properties of many-electron systems. Due to the
nonlinearity of the Kohn-Sham equations, the general
self-consistence searching method involves iterations with alternate
solving of the Poisson and Schr\"{o}dinger equations. One of
problems of such an approach is that the charge distribution renewed
by means of the Schr\"{o}dinger equation solution does not conform
to boundary conditions of Poisson equation for Coulomb potential.
The resulting instability or even divergence of iterations manifests
itself most appreciably in the case of infinitely extended systems.
The published attempts to deal with this problem are reduced in fact
to abandoning the original iterative method and replacing it with
some approximate calculation scheme, which is usually semi-empirical
and does not permit to evaluate the extent of deviation from the
exact solution. In this work, we realize the iterative scheme of
solving the Kohn-Sham equations for extended systems with
inhomogeneous electron gas, which is based on eliminating the
long-range character of Coulomb interaction as the cause of tight
coupling between charge distribution and boundary conditions. The
suggested algorithm is employed to calculate energy spectrum,
self-consistent potential, and electrostatic capacitance of the
semi-infinite degenerate electron gas bounded by infinitely high
barrier, as well as the work function and surface energy of simple
metals in the model with homogeneous distribution of positive
background. The difference between self-consistent Hartree solutions
and those taking into account the exchange-correlation interaction
is analyzed. The comparison with results previously published in the
literature is carried out. The case study of the metal-semiconductor
tunnel contact shows this method being applied to an infinitely
extended system where the steady-state current can flow.

\end{abstract}

\section{Introduction}\label{Sec-Intro}
 In the density functional approach, the set of equations describing
the inhomogeneous electron gas in the Kohn-Sham approximation should
be satisfied by the self-consistent distribution of electron density
$N({\mathbf r})$ and Coulomb potential $U({\mathbf r})$
\cite{K-V83}. Due to the essential nonlinearity of these equations,
the only general method of self-consistent solution involves
iterations using alternately the Poisson equation for potential and
the set of Schr\"{o}dinger equations for single-particle wave
functions in the effective potential $U_{\mathrm{eff}}({\mathbf r})=
U({\mathbf r}) + U_{\mathrm{xc}}({\mathbf r})$, where
$U_{\mathrm{xc}}$ is the exchange-correlation potential in the local
density approximation. A well-known problem of such an approach lies
in the necessity of taking into account the boundary conditions of
the Poisson equation, which impose certain requirements in terms of
integral relations on the charge distribution in the right-hand side
of this equation \cite{ASH06}. From the physical point of view,
these integral relations specify, for example, the total charge of
the electron system, if the boundary conditions are imposed on the
electric field, or the total dipole moment of the system, if the
values of potential on the boundaries of an infinite region are
fixed.

However, until the self-consistency is reached, the distribution of
electron density found by means of the Schr\"{o}dinger equation at
each step of iteration process turns out, as a rule, to be
incompatible with the Poisson boundary conditions specified.
Consequently, as shown in \cite{ASH06}, in the case of infinitely
extended systems it is either not possible to build the next step
solution at all, or the iteration process lacks convergence as
reported, for example, in \cite{L-K70PRB4555,Mann-Niem75PRB4012}. In
such a situation, some authors resort to replacing the process of
solving the Poisson equation with some kind of variation scheme
relative to numerical parameters that specify the functional form
chosen to approximate the potential or charge distribution. This is
discussed in more detail in subsections \ref{Subsec_self-levels} and
\ref{subsec_W simple met}.

 To manage this difficulty there are several artificial techniques
suggested, all of them changing the parameters of the system in
question, for example, the compensating background charge density
\cite{Liebsch97}. However, the existence and uniqueness of the
solution of self-consistent field equations follows from their being
obtained as Euler-Lagrange equations for a variational problem of
minimizing the total energy of non-degenerate ground state (see,for
example, \cite{K-V83}). In such a case the single-particle orbital
wave functions and Coulomb potential are varied at fixed boundary
conditions, while the parameters of the system remain constant. For
varying the system parameters over the iteration process one needs a
proof of existence and uniqueness of solution, which is now absent.

 Another approach suggested in \cite{Mann-Niem75PRB4012,Niem-JPhF77}
finds the electrostatic potential $U({\mathbf r})$ by means of a
linear integral equation of the second kind, to which the Poisson
equation is artificially reduced. In so doing, the necessity of
special measures to maintain compatibility of the Poisson boundary
conditions with charge distribution obtained at each iteration is
allegedly eliminated (see \cite{Niem-JPhF77}, Appendix). However it
is clear that the exact solution of integral equation, which is
equivalent to the original differential one, does not exist either,
if some of the compatibility conditions fail to hold. Apparently for
this reason, the said method, even if successful, was complemented
at each iteration  by some unclear renormalization of electron
density obtained by means of the Schr\"{o}dinger equation to fulfil
the neutrality condition before solving the Poisson equation
\cite{Mann-Niem75PRB4012}. One more drawback of this approach
appears to be, in fact, the conservation of screening term in the
region without electrons but with nonzero potential, as it is the
case in the surface problems.

 In this work we realize the iteration algorithm suggested earlier
and described in \cite{ASH06}, which is applicable to solving the
Kohn-Sham equations for the equilibrium inhomogeneous electron gas
and satisfies the boundary conditions for the Poisson equation at
any shape of electron distribution induced by the effective
potential of the previous approximation (section 2). The algorithm
peculiarities caused by discontinuity of quasiclassical expression
for the induced electron density with exchange-correlation potential
taken into account are pointed out (section
\ref{SSec-Algorithm/Applicat}). The convergence of the algorithm is
demonstrated by example of self-consistent calculations in a model
with homogeneous compensating background while studying such
problems as: 1) self-induced bound states near the surface of
barrier-bounded electron gas (section \ref{Subsec_self-levels}), 2)
parallel-plate capacitor with real distribution of screening charge
accounting for Friedel oscillations of electron density (section
\ref{SSec C-limit}), 3) work function and surface energy of simple
metals (section \ref{Sec Work function & surf energy}). The example
of a metal-semiconductor tunneling contact with Schottky barrier is
used to demonstrate the method's applicability to the systems where
the current flow is possible (section \ref{Sec-Schottky}). Some of
the results were preliminarily presented in \cite{Sh-P02} and
reported to the Russian conferences on semiconductor physics in 2001
- 2007.

\section {The base equations and iteration algorithm}\label{Sec-Algotithm}
\subsection{General formulation}\label{SSec-Algorithm/General}
 In the Kohn-Sham approximation, the gas of interacting electrons at
temperature $T=0$ is described by the set of equations for
single-particle wave functions $\Psi_{\varepsilon}$
\begin{equation}
\frac{1}{2}\nabla^2 \Psi_{\varepsilon}({\mathbf r}) + (\varepsilon
- U_{\mathrm{eff}}({\mathbf r}))\Psi_{\varepsilon}({\mathbf r})=0
\label{Schred-gen}
\end{equation}
with the effective potential
\begin{equation}
U_{\mathrm{eff}}({\mathbf r})= U({\mathbf r}) +
U_{\mathrm{xc}}({\mathbf r}). \label{Ueff-def}
\end{equation}
 Here
$\varepsilon$ is the energy eigenvalue of a single-particle state,
and $U$ is Coulomb potential energy of the electron described by the
Poisson equation
\begin{equation}
\nabla^2 U = 4 \pi (N_{+}({\mathbf r}) - N({\mathbf r})),
\label{Poisson-gen}
\end{equation}
where $N_{+}({\mathbf r})$ is positive charge density, $N({\mathbf
r})$ is electron density. Everywhere, unless otherwise specified,
the atomic units $|e|=m=\hbar=1$ are used based on charge $e$ and
mass $m$ of the free electron, when the unit of distance is Bohr
radius $a_{B}$, and the unit of energy is Hartree $Ha=e^{2}/a_{B}$.

 The exchange-correlation potential in equation (\ref{Ueff-def}) is taken in the
local approximation of density functional
\begin{equation}
U_{\mathrm{xc}}({\mathbf r}) =\left.
\frac{d[N\varepsilon_{\mathrm{xc}}(N)]}{dN}\right|_{N=N({\mathbf
r})}, \label{Pot_xc-def}
\end{equation}
where $\varepsilon_{\mathrm{xc}} = \varepsilon_{\mathrm{x}} +
\varepsilon_{\mathrm{c}}$ is the sum of exchange and correlation
energy per particle.

The electron density is expressed through wave functions as
\begin{equation}
N({\mathbf r})=2\!\!\int\limits_{\varepsilon \leq \varepsilon
_{\mathrm{F}}}\!\!\mathfrak{D}\left\{ \varepsilon \right\} \left|
\Psi _{\varepsilon }(\mathbf r)\right| ^{2}, \label{N(r)-def}
\end{equation}
where factor $2$ takes into account the spin degeneracy of
single-particle states, and the integration on differential spectral
measure $\mathfrak{D}\left\{ \varepsilon \right\}$ of Hamiltonian
$\hat{H}=-\frac{1}{2}\nabla ^{2}+U_{\mathrm{eff}}(\mathbf r)$ is
made over all occupied states with energy $\varepsilon$ not
exceeding the Fermi energy $\varepsilon _{\mathrm{F}}$.

As pointed out in the Introduction, the boundary conditions for the
Poisson equation impose certain requirements on the spatial
distribution of electron density, and the result of integration in
Eq. (\ref{N(r)-def}) does not necessarily comply with them until the
self-consistency is reached. However, if we conceive of the total
electron density as the sum
\begin{equation}
N({\mathbf r})=N_{\mathrm{ind}}(U(\mathbf r)) +
N_{\mathrm{qu}}({\mathbf r}), \label{N_ind-intro}
\end{equation}
where the induced density $N_{\mathrm{ind}}$ depends explicitly on
the unknown potential, the incompatibility problem of boundary
conditions with the right-hand side of Poisson equation does not
arise because the long-range Coulomb interaction is replaced with
the screened interaction of finite range. In the Kohn-Sham theory,
the electron density and effective potential can be coupled with an
approximate local relation similar to that used for the classical
ideal equilibrium Fermi gas
\begin{equation}
N_{\mathrm{ind}}({\mathbf r}) = \frac{2^{3/2}}{3\pi
^{2}}(\varepsilon _{\mathrm{F}} - U_{\mathrm{eff}}({\mathbf
r}))^{3/2}. \label{N_ind-def}
\end{equation}
Formula (\ref{N_ind-def}) can be obtained from the solutions of
Schr\"{o}dinger equation (\ref{Schred-gen}) for wave functions of
continuous spectrum in quasiclassical approximation by substituting
them into Eq. (\ref{N(r)-def}) and averaging the resulting electron
density over a scale larger than the Fermi wavelength $\lambda
_{\mathrm{F}}$. The function $N_{\mathrm{qu}}({\mathbf r})$ with
$N_{\mathrm{ind}}({\mathbf r})$ of Eq. (\ref{N_ind-def}) is a
quantum correction to the quasiclassical electron distribution. Note
that the representation of total density as in Eq.
(\ref{N_ind-intro}) is always exact regardless of using the
approximate formula (\ref{N_ind-def}) for $N_{\mathrm{ind}}({\mathbf
r})$.

In expression (\ref{Ueff-def}), $U_{\rm xc}=U_{\rm x}+U_{\rm c}$.
The exchange potential for electron gas in the local approximation
is given by the well-known expression
\begin {equation}
U_{\rm x}(\mathbf{r})=-\left(  \frac{3}{\pi}\right)  ^{1/3}N^{1/3}(\mathbf{r}%
)=-\left(  \frac{3}{2\pi}\right)^{2/3}\frac{1}{r_{s}(\mathbf{r})}.
\label{U_x-def}
\end {equation}
The correlation potential, according to Eq. (\ref{Pot_xc-def}), is
defined through the correlation energy per particle
$\varepsilon_{\mathrm{c}}$. We assume the latter as
\begin{equation}
\varepsilon_{\mathrm{c}}=-\frac{0.44}{r_{\mathrm{s}} + 11.5}.
\label{E_c-def}
\end{equation}
Here $r_{\mathrm{s}}$ is the local Wigner-Seitz radius defined as
\begin{equation} \label{r_s-def}
4\pi r_{\mathrm{s}}^{3}(\mathbf r)/3=N^{-1}(\mathbf r).
\end{equation}
For the following analysis it is convenient to introduce
$R_{\mathrm{s}}=r_{\mathrm{s}}(\infty)$ as the value of this
parameter at infinity, where we always take the local neutrality
condition $\lim_{|\textbf{r}| \to \infty}(N-N_+) \to 0$ to be
fulfilled.

Having calculated the derivative (\ref{Pot_xc-def}) with correlation
energy given by Eq. (\ref{E_c-def}), one can write the correlation
potential as
\begin{equation}
U_{\mathrm{c}}(\mathbf{r})=-\frac{22}{75}\frac{8r_{s}(\mathbf{r})+69}{\left[
2r_{s}(\mathbf{r})+23\right]  ^{2}}.\label{Uc-expli}%
\end{equation}

The formula (\ref{E_c-def}) for correlation energy is a well-known
Wigner expression with a different parameter $11.5$ instead of $7.8$
in denominator. The reasons for using the modified Wigner formula
will be discussed in more detail elsewhere. Here we only note that
such a choice of $\varepsilon_{\mathrm{c}}$, on the one hand,
describes better the $R_{\mathrm{s}}$ dependence of cohesive energy
of simple metals represented in Table 8 of the book \cite{Pines-63}.
On the other hand, the value $R_{\mathrm{s}}=5.63$ accepted in the
literature for \texttt{Cs} turns out less than the critical one
$R_{\mathrm{s\, c}}=5.64$, which characterizes the stability of
homogeneous electron gas in the jellium model with respect to
spatially inhomogeneous perturbations. The impossibility to use the
formula (\ref{N_ind-def}) for induced charge in the case of
$R_{\mathrm{s}} \geq R_{\mathrm{s \,c}}$ would make the suggested
algorithm inapplicable, so the choice of correlation energy in the
form (\ref{E_c-def}) provides the existence of cesium metal in the
framework of calculation technique under development. There are also
other expressions suggested in the literature for correlation energy
in the local density approximation LDA (see, for example, discussion
in \cite{K-V83,Liebsch97}), but for the purposes of this work aimed
at demonstrating in detail the application of the new algorithm to a
number of classical problems, the specific form of correlation
potential is not so important, because the relevant modifications to
the expression for $U_{\mathrm{c}}$ lead to only quantitative
changes in calculation results, and do not create fundamental
difficulties for realization of iteration scheme suggested.

The separation of induced charge brings the Poisson equation
 (\ref{Poisson-gen}) to the form
\begin{equation}
\nabla^2 U + 4\pi N_{\mathrm{ind}}(U) = 4\pi(N_+({\mathbf r}) -%
N_{\mathrm{qu}}({\mathbf r})). \label{Poiss-N_ind}
\end{equation}
The quantum correction to electron density $N_{\mathrm{qu}}({\mathbf
r})$ introduced here by means of Eqs.
(\ref{N(r)-def})-(\ref{N_ind-def}) is defined as
\begin{equation}
N_{\mathrm{qu}}({\mathbf r}) =2\int\limits_{\varepsilon \leq
\varepsilon _{\mathrm{F}}}\mathfrak{D}\left\{ \varepsilon \right\}
\left| \Psi _{\varepsilon }(\mathbf r)\right| ^{2} -
N_{\mathrm{ind}}({\mathbf r}). \label{N_qu-def}
\end{equation}
The algorithm of consecutive steps $i = 0,1,2,..$ for iterative
solving of equations (\ref{Schred-gen}) - (\ref{N(r)-def}) with
equation (\ref{Poisson-gen}) changed to (\ref{Poiss-N_ind}) can be
now formulated as follows:
\begin{align}
\intertext{1. Neglect the small-scale quantum variations of electron
density}
& N_{\mathrm{qu}}^0=0; \label{N_qu-init}\\
\intertext{2. Solve the nonlinear Poisson equation with a known
right-hand side} & \nabla^{2} U^{i}({\mathbf r}) + 4\pi
N_{\mathrm{ind}}(U^{i},N_{\mathrm{qu}}^i)=4\pi \left(N_{+}({\mathbf
r}) -
N_{\mathrm{qu}}^i({\mathbf r})\right); \label{Poisson-i}\\
\intertext{3. Find the total electron density self-consistent with
$U$, and effective potential} & N_{\mathrm{s}}^i({\mathbf r}) =
N_{\mathrm{ind}}^i({\mathbf r}) + N_{\mathrm{qu}}^i({\mathbf
r});\quad U_{\mathrm{eff}}^{i}({\mathbf r}) = U^{i}({\mathbf
r})+U_{\mathrm{xc}}
\left(N_{\mathrm{s}}^i({\mathbf r})\right); \label{N^i,U^i-cycle}\\
\intertext{4. Find the eigenfunctions of the single-particle
Kohn-Sham Hamiltonian}
&\frac{1}{2}\nabla^{2}\Psi_{\varepsilon}^{i}({\mathbf r}) + \left(
\varepsilon - U_{\mathrm{eff}}^{i}({\mathbf r}) \right)
\Psi_{\varepsilon}^{i}({\mathbf r})=0;
\label{Schred^i-cycle}\\
\intertext{5. Find anew the total electron density, using the
eigenvalues and eigenfunctions obtained} & N^{i}({\mathbf r}) =
2\!\!\int\limits_{\varepsilon \leq \varepsilon
_{\mathrm{F}}}\!\!\mathfrak{D}\left\{ \varepsilon \right\} \left|
\Psi _{\varepsilon }^{i}(\mathbf r)\right| ^{2};
\label{Ni_gen def}\\
\intertext{6. Calculate the quantum correction for the next
iteration step by means of the new total electron density}
&N_{\mathrm{qu}}^{i+1}({\mathbf r}) = N^{i}({\mathbf r}) -
N_{\mathrm{ind}}(U_{\mathrm{eff}}^{i}({\mathbf r}));
\label{N_qu-next}
\end{align}
7. Return to solving the Poisson equation with the new right-hand
side.

Index $i$ here denotes the iteration number. It is of importance
that the Coulomb potential $U^{i}({\mathbf r})$ and the induced
electron density $N_{\mathrm{ind}}^i({\mathbf r})$ are found
self-consistently at a fixed quantum density
$N_{\mathrm{qu}}^{i}({\mathbf r})$ in the process of solving the
Poisson equation \eqref{Poisson-i}.

The convergence criterion for iteration process is the value of
$\delta$, which limits the maximal deviation of the electron density
$N_{\mathrm{s}}^{i}(\mathbf r)$ obtained after solving the Poisson
equation and self-consistent with the potential $U^{i}(\mathbf r)$,
from the density $N^{i}(\mathbf r)$ found by means of the
Schr\"{o}dinger equation,
\begin{equation} \label{Nevyazka_N-gen}
\max_{\mathbf{r}} \left| N_{\mathrm{s}}^{i}({\mathbf
r})-N^{i}({\mathbf r}) \right|\left/ N(\infty) \right. \leq \delta.
\end{equation}
It is assumed here that electron density at infinity is determined
by the local neutrality condition and does not change through the
iteration process.

In the course of solving the self-consistent Poisson equation
(\ref{Poisson-i}), the induced density
$$N_{\mathrm{ind}}^i=N_{\mathrm{ind}}(U^{i},N_{\mathrm{qu}}^i)$$
should be calculated as an implicit function of unknown potential
 $U^{i}$ at a known quantum correction
$N_{\mathrm{qu}}^{i}$ according to formulae (\ref{Ueff-def}) and
(\ref{N_ind-def}). It is implicit because the exchange-correlation
potential in Eq. (\ref{N_ind-def}) depends on the total electron
density $N_{\mathrm{ind}}^{i}+N_{\mathrm{qu}}^{i}.$

One should note that representing the induced charge as Eq.
(\ref{N_ind-def}) requires the condition
\begin{equation}\label{Stability cond}
\frac{dN_{\mathrm{ind}}}{dU} < 0.
\end{equation} to be fulfilled.

Otherwise, the square of the linear screening length in the equation
(\ref{Poisson-i}) linearized for $U$ turns negative, and the
response to small spatial perturbations of electron density is no
more localized due to oscillating behavior of solutions of the
self-consistent Poisson equation. The inequality (\ref{Stability
cond}) is generally violated in the regions where electron density
is small and, accordingly, the relative role of exchange-correlation
potential increases. In such regions of instability, the induced
electron density cannot be represented with Eq. (\ref{N_ind-def}),
so we take $N_{\mathrm{ind}}^i({\mathbf r})\equiv 0$. This situation
occurs, for example, near the surface of electron systems at the
distribution tails where the electron density is close to zero.

To conclude this section, there are some remarks on terminology used
further on. The result of self-consistent calculation by means of
the complete set of equations \eqref{N_qu-init}-\eqref{N_qu-next}
with expression \eqref{N_ind-def} for the induced electron density,
exchange potential as \eqref{U_x-def} and correlation potential
\eqref{Uc-expli} is called the exact or the self-consistent
solution. Solving the same set of equations, but ignoring $U_{\rm
xc}$, one obtains the self-consistent solution in the Hartree
approximation. The self-consistent solution of the Poisson equation
\eqref{Poiss-N_ind} with the induced density \eqref{N_ind-def},
ignoring $U_{\rm xc}$, and at $N_{\rm qu}=0$ is the Thomas-Fermi
approximation. If the exchange potential $U_{\rm x}$ is taken into
account in the latter scheme, it becomes the Thomas-Fermi-Dirac
approximation. Using the total effective potential in the local
density approximation with the induced charge formula
\eqref{N_ind-def} is sometimes called the Thomas-Fermi-Dirac-Gombas
approximation (see, for example, \cite{Gombas-49}, Sec. 11).

\subsection{Peculiarities of the algorithm application}
\label{SSec-Algorithm/Applicat}

The calculation technique using the induced electron density is
clarified more specifically in further sections, where the suggested
iteration algorithm is realized in calculating the quantum
correction to the capacitance of barrier structures and in finding
the work function and surface energy of simple metals in the model
with homogeneous positive background. At the same time, the
homogeneity of background allows to get still further in the
analytical transformation of the general equations. Let us consider
the semiinfinite electron gas assuming that the ion charge forms a
homogeneous positive background of density $N_+$ given by
\begin{equation}\label{N+semi}
N_+(z) = \left\{
\begin{aligned}
& N_{+}, &z \geq z_+,\\
& 0, &z < z_+.\\
\end{aligned}
\right.
\end{equation}
In this model, the effective potential in the Schr\"{o}dinger
equation (\ref{Schred^i-cycle}) depends solely on the coordinate
$z$. Such system can represent, for example, a metal surface with
self-consistent barrier or a semiconductor structure bounded by a
potential barrier.

The expression for induced charge in the form (\ref{N_ind-def}) is
asymptotically correct (with respect to quasiclassicality parameter,
which is always small in the systems in question at $z \to \infty$)
in describing the self-consistent reaction of electrons to the
change in the long-wave part of Coulomb potential, which is what is
necessary to eliminate the infinite range of direct Coulomb
interaction of electrons. However, as noted before, it fails in the
regions of small electron density where the condition
\eqref{Stability cond} does not hold. In view of the above, the
self-consistent Poisson equation can be written as
\begin{equation} \label{Poisson-crit point}
\begin{aligned}
& \frac{d^{2}u^{i}}{d\zeta^{2}}+c_{n}n_{\rm ind}(u^{i},n_{\rm qu}^{i})%
 =c_{n}(\theta(\zeta-\zeta_+)-n_{\rm qu}^{i}(\zeta)),\,\,\,\,& \zeta
>\zeta_{\rm c}^i\\
& \frac{d^{2}u^{i}}{d\zeta^{2}}
=c_{n}(\theta(\zeta-\zeta_+)-n^{i}(\zeta)),\,\,\,\,& \zeta
\leq\zeta_{\rm c}^i
\end{aligned}
\end{equation}
where $c_{n}=(8/3\pi)\left( 4/9\pi \right)^{1/3}R_s$, and
$\theta(\zeta)$ is the Heavyside theta function. We introduced the
dimensionless variables $\zeta = k_{\rm F}z$, $\zeta_+ = k_{\rm
F}z_+$, $n(\zeta) = N(z)/N_+$, $u(\zeta) = U(z)/ \varepsilon_{\rm
F}^0$, where $k_{\rm F}=(3\pi^2N_+)^{1/3}$, $\varepsilon_{\rm
F}^0=k_{\rm F}^2/2$. The boundary conditions must force the electric
field $du/d\zeta$ to be zero at infinity, to ensure the potential
finiteness, $u(\infty)=const$, and local neutrality of the electron
gas, $n(\infty)=1$.

Eq. (\ref{Poisson-crit point}) assumes the existence of a single
critical point $\zeta_{\rm c}$ that separates the region with the
induced density $n_{\rm ind}$ expressed as \eqref{N_ind-def}, from
the region, where such expression is invalid according to criterion
\eqref{Stability cond}. When calculating the systems that meet the
condition $R_{\rm s}<R_{\rm sc}$, it is always possible to introduce
$n_{\rm ind}$ in the region $\zeta_{\rm c}^{i}< \zeta < \infty$. The
density $n_{\rm qu}^{i}$ turns out definite at the same region.
Actually the problem is always solved on the finite interval of real
axis, $0 \leq \zeta \leq \zeta_{\rm max}$. The value $\zeta_{+} \geq
0$ denotes the possible shift of the positive background starting
point from zero. Values $\zeta_{+}$ and $\zeta_{\rm max}$ are
assumed to be taken so large that the transfer of boundary
conditions to the points $\zeta=0$ and $\zeta_{\rm max}$ from
$\zeta=-\infty$ and $\zeta=\infty$ does not affect the result within
the chosen accuracy. This condition is easily checked by
calculations at several consecutively increasing values of
$\zeta_{+}$ and $\zeta_{\rm max}$.

At the point $\zeta_{\rm c}$ the right-hand and left-hand solutions
are matched together so that potential $u$ is continuous along with
its first derivative. The critical point $\zeta_{\rm c}$ is defined
as the point where inequality(\ref{Stability cond}) reverses its
sign, that is, the derivative $dn_{\mathrm{ind}}/du$ turns positive.
As long as with reversing sign the derivative of multivalued
implicit function $n_{\mathrm{ind}}(u)$ passes through infinity, it
is more convenient, in the manner of the density functional
approach, to analyze the inverse single-valued function
$u(n_{\mathrm{ind}})$, whose derivative reverses its sign, passing
through zero. This function can be found from formula
(\ref{N_ind-def}) and in the dimensionless variables introduced has
the form
\begin{equation}\label{u->n_ind}
    u\left(n_{\mathrm{ind}}\right)=\mu -n_{\mathrm{ind}}^{2/3}-u_{\mathrm{xc}}\left(%
    n_{\mathrm{ind}}+n_{\mathrm{qu}}\right),
\end{equation}
where $u_{\mathrm{xc}}=U_{\mathrm{xc}}/\varepsilon_{\mathrm{F}}^0$,
and $\mu=1-u_{\mathrm{xc}}(\infty)$ is the dimensionless Fermi
energy of electrons accounting for the exchange-correlation
interaction.

Unlike the analysis of validity conditions for the formula of
induced density in homogeneous electron gas carried out in
\cite{ASH-Zai_SSC76}, which provides, along with (\ref{E_c-def}),
the estimate of critical density as $R_{\mathrm{sc}}=5.64$, in
present case the dependence $n_{\mathrm{ind}}(u)$ is also affected
by the quantum correction $n_{\mathrm{qu}}$ to the density. This
influence is caused by the exchange-correlation potential in the
local density approximation being dependent on the total electron
density. For that reason, the analysis should be carried out all
over again on the base of the expression
\begin{equation}\label{du/dn_ind}
    \frac{du}{dn_{\mathrm{ind}}}=-\frac{2}{3}n_{\mathrm{ind}}^{-1/3}-\frac
{d}{dn_{\mathrm{ind}}} \left[
u_{\mathrm{xc}}\left(n_{\mathrm{ind}}+n_{\mathrm{qu}}\right)
\right]
\end{equation}

The result needed for qualitative understanding can be derived
analytically, the correlation energy being neglected. Its
consideration does not change the general picture, for even at the
least acceptable value of density of the homogeneous electron gas at
the infinity, corresponding to the highest acceptable value
$R_{\mathrm{sc}}=5.64$, the ratio
$u_c(R_{\mathrm{sc}})/u_x(R_{\mathrm{sc}})=0.26$, and the derivative
ratio is even less, ($du_{\mathrm{c}}/dn)/(du_{\mathrm{x}}/dn)\simeq
0.07$. With a rise in density, these estimates still improve in the
sense favorable to our purposes. At the same time, in the course of
numerical realization of the algorithm the condition
$du/dn_{\mathrm{ind}}\leq 0$ is easily checked without any
simplification by the numerically known right-hand side of formula
\eqref{du/dn_ind}.

All the above considered, the inequality that determines the
validity range of quasiclassical expression (\ref{N_ind-def}) is
conveniently written as
\begin{equation}\label{x-inequality}
    -\frac{2}{3}n_{\mathrm{ind}}^{-1/3}+\frac{1}{3}c_{\mathrm{x}}R_{\mathrm{s}%
}\left(  n_{\mathrm{ind}}+n_{\mathrm{qu}}\right)  ^{-2/3}\leq0,
\,\,\,c_{\mathrm{x}}=2\left(2/3\pi ^2 \right)^{2/3}
\end{equation}
The derivative of the exchange potential (\ref{U_x-def}) is
expressed here in our dimensionless variables. On the basis of
(\ref{x-inequality}) one can obtain the equation for critical values
$n_{\rm c}$ of total density $n=n_{\mathrm{ind}}+n_{\mathrm{qu}}$,
which correspond to changing in sign of the derivative
$du/dn_{\mathrm{ind}}$,
\begin{equation}\label{nc-equ}
    n_{\mathrm{c}}^{2}-\beta n_{\mathrm{c}}+\beta n_{\mathrm{qu}}=0,\,\,\,\,\beta
=(c_{\mathrm{x}}R_{\mathrm{s}})^{3}/8.
\end{equation}
Its solutions should satisfy the condition of non-negative total
density, $n\geq 0$, and induced density, $n_{\mathrm{ind}}\geq 0$,
by virtue of their definition by formulae (\ref{N(r)-def}) and
(\ref{N_ind-def}). The value of the quantum correction
$n_{\mathrm{qu}}$ can be of any sign, because it is defined by
formula (\ref{N_ind-intro}) as a difference of two expressions for
electron density found from the exact and approximate solutions of
the Schr\"{o}dinger equation.

The roots of equation (\ref{nc-equ}) are given by
\begin{equation}\label{nc-roots}
    n_{\mathrm{c1,2}}=\frac{\beta}{2}\left(  1\pm\sqrt{1-\frac{4}{\beta
}n_{\mathrm{qu}}}\right).
\end{equation}
It is clear that with $n_{\mathrm{qu}}\leq 0$ only one root
\begin{equation}\label{nc1}
    n_{\mathrm{c1}}=\frac{\beta}{2}\left( 1+\sqrt{1-\frac{4}{\beta
}n_{\mathrm{qu}}}\right),
\end{equation}
satisfies the stated condition of non-negativity $n_{\mathrm{c}}$
and $n^{\mathrm{(c})}_{\mathrm{ind}}=n_{\mathrm{c}}-n_{\mathrm{qu}}$.
In the case of $n_{\mathrm{qu}}=0$, formula (\ref{nc1}) gives the
value of $n_{\mathrm{c}}$ for the homogeneous electron gas allowing
for only the exchange potential, which after switching to atomic
units corresponds to $R_{\mathrm{sc}}=6.02$. At $n_{\mathrm{qu}}<0$,
the critical density is even higher. Taking into account the
correlation potential also increases the estimate for
$n_{\mathrm{c}}$, adding a positive term to the left-hand side of
inequality (\ref{x-inequality}). In the case of a homogeneous gas
with correlation energy as (\ref{E_c-def}), this shifts the critical
value towards $R_{\mathrm{sc}}=5.64$.

While negative values of $n_{\mathrm{qu}}$ do not change
qualitatively the dependence of induced density $n_{\mathrm{ind}}$
on the average Coulomb potential $u$ compared to the homogeneous
case, leaving only one value of critical density, there is quite
another picture at $n_{\mathrm{qu}}>0$. Both roots given by formula
(\ref{nc-roots}) have physical meaning and ensure positivity of both
total density $n_{\mathrm{c}}$, which is evident, and the induced
one $n_{\mathrm{c}}-n_{\mathrm{qu}}$, which can be easily checked.
It can also be seen that there is the limitation
$n_{\mathrm{qu}}\leq\beta/4$ on the value of quantum correction to
provide the existence of the roots. If $n_{\mathrm{qu}}$ exceeds
this limit, the Coulomb potential $u$ becomes a steadily decreasing
function of $n_{\mathrm{ind}}$ in the total range of physically
acceptable values, and the applicability condition (\ref{Stability
cond}) for the induced density formula (\ref{N_ind-def}) holds at
any values of $u$ and $n$.

It should be emphasized that the conditions for $n_{\mathrm{c}}$
caused by inequality (\ref{x-inequality}) do not coincide with the
condition of non-negativity of the radicand in formula
(\ref{N_ind-def}), coming into action earlier. It is due to this
circumstance caused by negativity of $u_{\rm xc}$ and its dependence
on $n$ that formula (\ref{N_ind-def}) breaks down earlier with
growing potential $u$, than $n_{\rm ind}$ turns to zero. So at the
increasing $u$ or decreasing $n$, the induced charge drops to zero
by a jump from some finite value in the course of iterations,
creating discontinuities of the first kind in the distribution of
total electron density and the effective potential, as can be seen
in Fig. \ref{fig1-n(z)-jump}. After the self-consistency is
achieved, this discontinuity naturally turns out to be less than
accuracy specified, i. e. the loop termination criterion
(\ref{Nevyazka_N-gen}). Nevertheless, the treatment of this
discontinuity when solving the Schr\"{o}dinger and Poisson equations
turned out an essential factor touching upon the effectiveness of
the whole algorithm, what will be discussed further.

The estimate of the value of discontinuities in the density and
effective potential can be made accurately, neglecting
$n_{\mathrm{qu}}$. It follows from the results of such analysis that
the value of density discontinuity equals
$n_{\mathrm{ind}}^{\mathrm{(c)}}$, taking on the value
$\left(R_{\mathrm{s}}/R_{\mathrm{sc}}\right) ^{3}$ at the initial
cycle of iterations. One can see that with a growing non-ideality of
electron gas and $R_{\mathrm{s}}$ closing to $R_{\mathrm{sc}}$ the
dimensionless density jump tends to unity and becomes a more and
more strong perturbation.

The discontinuity of effective potential equals
$u_{\mathrm{xc}}\left[ n_{\mathrm{ind}}^{\mathrm{(c)}}\right]$,
because the Coulomb potential is continuous everywhere. In the
atomic units, this discontinuity has an universal value dependent
only on the expression chosen for correlation energy and ranges from
$0.1\mathsf{Ha}$ with just the exchange potential taken into account
to $0.15\mathsf{Ha}$ with the correlation potential added according
to the standard Wigner formula \cite{ASH06}. In the dimensionless
units, however, the discontinuity of effective potential also grows
considerably with the increase of electron non-ideality due to
decreasing $\varepsilon_{\rm F}^0$.

Now let us write down the algorithm of iterative solution for Eqs.
(\ref{N_qu-init})-(\ref{N_qu-next}) as applied to the case of
semi-infinite many-electron systems with homogeneous positive
background considered in this section, using the dimensionless
variables introduced and duly commenting every step taken.

The initial value of the quantum correction to electron density and
the total density itself are taken to be zero, and the initial
position of critical point is taken beyond the boundary of positive
background:
\begin{equation}\label{nq-init}
n_{\mathrm{qu}}^0 (\zeta)\equiv 0, \,\,\, n^0(\zeta)\equiv 0, \,\,\,
\zeta^{0}_{\rm c}=\zeta_{+}-1.
\end{equation}
Solving the Poisson equation \eqref{Poisson-crit point}, we find the
Coulomb potential $u^i$ satisfying the specified boundary conditions
and a new total electron density $n_{\mathrm{s}}^i$, self-consistent
with it,
\begin{equation}\label{ns-def}
    n_{\mathrm{s}}^i=n_{\mathrm{ind}}^i +n_{\mathrm{qu}}^i,\,\zeta
    >\zeta_{\mathrm{c}}^{i},\,\,\,\,\,\
     n_{\mathrm{s}}^i=n^i,\,  \zeta \leq \zeta_{\mathrm{c}}^{i},\,\,\,\,\, i=0,1,...
\end{equation}
In the process, a check for condition $dn_{\mathrm{ind}}^i/du^i < 0$
can show that critical point should be relocated to a new position
$\zeta_{\rm c}^{i+1}$. In the presence of functions $n_{\rm
s}^i(\zeta)$ and $u^i(\zeta)$ being numerically known, the simplest
way of this check is to apply the formula (\ref{du/dn_ind}), where
$n_{\mathrm{ind}}^i$ is replaced with its expression in terms of
potential $u^i$ and definition (\ref{N_ind-def}) is rewritten in
dimensionless variables:
\begin{equation}\label{n_ind-u^i}
n_{\mathrm{ind}}^i=\left(\mu-u^i-u_{\mathrm{xc}}(n_{\mathrm{s}}^i)
\right)^{3/2}.
\end{equation}
If critical point is shifted to the right, one should redefine
$n_{\mathrm{s}}^i$, replacing $\zeta_{\mathrm{c}}^{i}$ with
$\zeta_{\mathrm{c}}^{i+1}$ in the procedure (\ref{ns-def}). Shift of
the critical point to the left leaves $n_{\mathrm{s}}^i$ unchanged,
though the new position of critical point defines the region of
induced density formation after the Schr\"{o}dinger equation has
been solved.

The case of the initial cycle, $i=0$, stands somewhat apart because
the direct numerical solution of the second-order differential
equation can be bypassed. Taking $\zeta_{\mathrm{c}}^{0}=0$, one
should begin the solution with Eq. \eqref{Poisson-crit point}. In
the absence of the quantum correction to density, when the induced
charge, nonlinear in $u$, does not depend explicitly on the
coordinate $\zeta$, the first integral of this equation can be found
in an explicit form. The solution is then found as numerical
quadrature of the first integral. In that solving one determines the
Coulomb potential $u$, the electron distribution self-consistent
therewith and reduced here just to $n_{\mathrm{ind}}^0$, and the
critical point position $\zeta_{\mathrm{c}}^{1}$. If the system as a
whole is sufficiently described by quasiclassical approximation,
such a solution may have a practical significance (see, for example,
formulae (35.70)-(35.73) in \cite{Gombas-49} and the calculation of
tunneling current through the Schottky barrier in
\cite{ASH-Zai_SSC76}).

With the solution of the Poisson equation known, one can form the
effective potential
$u_{\mathrm{eff}}^i=u^i+u_{\mathrm{xc}}^i(n_{\mathrm{s}}^i)$ that
enters into the Schr\"{o}dinger equation. In virtue of translation
invariance of the system under study in $(x,y)$ plane, the total
wave function can be written as
\begin{equation}\label{full w-function}
\Psi_{k,\mathbf{k}_{||}}\left(  \zeta,\mathbf{r}_{||}\right)
=\frac{1}{2\pi}e^{i\mathbf{k}_{||}\mathbf{r}_{||}}C\psi_{k}(\zeta),
\end{equation}
where vector $\mathbf{r}_{||}$ lies in the $(x,y)$ plane, and
constant $C$ is defined by the normalization requirement for
function $\psi_{k}$. The Schr\"{o}dinger equation for $\psi_{k}^{i}$
at the $i$-th iteration cycle is conveniently written as
\begin{equation}\label{schred^i}
   \frac{d^{2}}{d\zeta^{2}}\psi_{k}^{i}(\zeta)+\left(k^{2}-u_{\mathrm{eff}}^i(\zeta)%
   +u_{\mathrm{xc}}(\infty)\right)
\psi_{k}^{i}(\zeta)=0,
\end{equation}
where the dimensionless quantum number $k$ is normalized to $k_{\rm
F}$, and the scalar potential is calibrated by the condition
$u(\infty)=0$. An eigenvalue of $k$ specifies the behavior of
continuous spectrum eigenfunctions normalized to
$\delta(k-k^{\prime})$ according to \cite{ASH06} (see formulae
(24)-(25)) with the asymptotic form at infinity
\begin{equation}\label{psi-asympt}
\psi_{k}^{i}(\zeta) \rightarrow
\sqrt{\frac{2}{\pi}}\sin(k\zeta+\gamma_k), \,\,\,\, \zeta
\rightarrow \infty.
\end{equation}
The numerical solution $\tilde{\psi}_{k}(\zeta)$ of the Cauchy
problem for the homogeneous equation (\ref{schred^i}), found
accurate within an arbitrary factor, is normalized correctly, if its
asymptote at $\zeta \rightarrow \infty$ is as Eq.
\eqref{psi-asympt}. This is achieved multiplying the obtained
solution by a constant $A$ defined by relation
\begin{equation}\label{Eq amplit-def}
   A^2\left( \tilde{\psi}_{k}^2(\zeta_{\infty})+\tilde{\psi}^{\prime 2}_{k}(\zeta_{\infty}) \right)=2/\pi.
\end{equation}
Here $\zeta_{\infty}\leq \zeta_{max}$ is a sufficiently remote point
where potential $u_{\rm eff}(\zeta_{\infty})$ can be considered as
become a constant. The phase $\gamma_k$ can be found from the
relation
\begin{equation}\label{Eq phase-def}
\gamma_{k}=\arctan\left[  k\frac{\tilde{\psi}_{k}(\zeta_{\infty})}%
{\tilde{\psi}_{k}^{\prime}(\zeta_{\infty})}\right]  -\left(
k\zeta_{\infty }-l\pi\right),
\end{equation}
where $l=[k\zeta_{\infty }/\pi]$ is an integral part of the ratio.
With such a definition, the phase is invariant with respect to the
choice of $\zeta_{\infty}$ point.

To conclude the $i$-th iteration cycle, the total electron density
is calculated,
\begin{equation}\label{n^i-calc}
    n^{i}(\zeta)=3\int_{0}^{1}dk(1-k^{2})\left| \psi_{k}^{i}(\zeta) \right|^{2}
\end{equation}
as well as the quantum correction to the density for the next cycle
\begin{equation}\label{n_ind^(i+1)}
    n_{\mathrm{qu}}^{i+1}(\zeta)=n^{i}(\zeta)-n_{\mathrm{ind}}\left(
u_{\mathrm{eff}}^{i}(\zeta)\right),
\end{equation}
with the definition \eqref{n_ind-u^i} taken into account.

The Cauchy problem for the Schr\"{o}dinger equation was solved
numerically using the fourth-order implicit technique that is often
called the Numerov's method. The Numerov's algorithm does not
include directly the value of first derivative of the sought
solution specified at the initial point. Instead of this, one should
know the value of the sought function at the first point from the
boundary. To calculate this value with accuracy matching that of the
Numerov's method, we represented the function as a second-order
expansion in Taylor series. The necessary value of the second
derivative at the boundary was found by means of the Schr\"{o}dinger
equation itself. A similar technique was applied to pass the
discontinuities of effective potential. It has been checked within
the chosen accuracy that the result of such a procedure is in accord
with Runge-Kutta method commonly used to move one step away from the
point where the Cauchy boundary conditions are specified.

The Poisson equation was solved by the relaxation method
\cite{Raznost-schem77}, its idea being to solve the matching
nonstationary equation
\begin{equation}\label{poisson_tdpt}
    \frac{\partial u}{\partial t}-\left(  \frac{\partial^{2}u}{\partial\zeta^{2}%
}+c_{\mathrm{n}}n_{\mathrm{ind}}(u,n_{\mathrm{qu}})\right)  = -c_{\mathrm{n}}%
[\theta(\zeta-\zeta_{+})-n_{\mathrm{qu}}(\zeta)]
\end{equation}
instead of the stationary problem, until the solution stops changing
in time $t$ within the accuracy specified.

From the formal point of view, Eq.(\ref{poisson_tdpt}) is a
nonlinear diffusion equation, a lot of numerical schemes having been
developed for its solution. In present work we applied a scheme, in
which, after discretization \eqref{poisson_tdpt} in time with step
$\tau$ and in space with step $h$, one specifies an initial
potential distribution satisfying the boundary conditions, and time
evolution of the solution is monitored by solving the boundary value
problem at each time point. As this takes place, one need not to
obtain with a good accuracy the spatial distribution of potential
$u$ at each "time step", searching only for an accurate stationary
solution at long times when change in $u$ from step to step becomes
small. The time discretization was made with an implicit first-order
scheme, so, when forming the boundary value problem at a new time
point, the nonlinear term in the equation was estimated via the
induced density $n_{\mathrm{ind}}(u,n_{\mathrm{qu}})$ linearized in
the increment of potential at the time step $\tau$. In the course of
solving Eq. \eqref{poisson_tdpt}, the choice of $\tau$ value was
dictated, on the one hand, by considerations of numerical stability,
and on the other hand, by those of computation speed (the time of
reaching the stationary solution). In line with recommendations
given in the literature, the initial time step was chosen according
to relation $ \tau/h^2=0.1$.

Since the linearization of dependence $n_{\rm ind}(u)$ was used at
the movement from one time point to the next, the necessity could
arise in the course of solving the Poisson equation to redefine the
critical point to avoid the appearance of a region where $dn_{\rm
ind}(u,n_{\rm q})/du>0$. This was carried out similarly to the
procedure described after formula (\ref{n_ind-u^i}). As noted above,
the magnitude of ordinary jump of the electron density increases
with $R_{\rm s}$ and leads to more and more violent perturbations of
numerical solution. It was found (see details in Sec. \ref{Sec Work
function & surf energy}) that at $R_{\rm s} \lesssim R_{\rm sc}$ the
convergence of solution of the equation set (\ref{N_qu-init}) -
(\ref{N_qu-next}) deteriorates. It appears though to be rather a
consequence of some specific numerical algorithm realization, than a
drawback of the iteration scheme suggested.

\section {Properties of electron gas in the barrier structures}
\label{Sec_barrier structures}

In this section, we apply our method of solving the Kohn-Sham
equations to finding the self-consistent electron density and
effective potential in a barrier structure that consists of two
conductors with degenerate electron gas separated by an insulating
layer with dielectric permeability $\kappa_{\rm d}$. We assume that
z-axis is perpendicular to the layer plane, the conductors are
positioned at $z \leq -d/2$ and $z \geq d/2$ respectively, and the
insulating layer lies at $|z| < d/2$. It is assumed further on that
the neutralizing background in the conductors is homogeneous, and
the barrier is not impurity-doped, that is,

\begin{equation}\label{Eq. barrier N+}
    N_+(z) = \left\{
\begin{aligned}
& 0, &|z| < d/2,\\
& N_{+}, &|z|\geq d/2.
\end{aligned}
\right.
\end{equation}
The height of potential barrier formed by the insulator is taken to
be infinite, so that wave functions turn to zero at the
conductor-insulator interface.

\subsection{Self-induced energy levels in semiconductors}
\label{Subsec_self-levels}

In the case of infinitely high potential barrier, the electron
states localized in $z$ and the size-quantized two-dimensional
subbands can emerge even if inhomogeneities of positive background
$N_+(z)$ and applied voltage are absent. The existence of such
states was suggested for the first time in \cite{konst-shik}, where
the discrete energy levels in question were called "the plasma
levels". The appearance of localized states was attributed to the
electron wave functions turning to zero at the barrier interface.
Due to continuity of wave functions, the electron density close to
the barrier is small and inadequate to provide the local neutrality.
The uncompensated positive charge near the interface creates a
self-consistent potential well, wherein bound states can exist.
According to calculation carried out in \cite{konst-shik} for the
case of degenerate electron gas, the discrete levels should exist at
$R_{\rm s} \le 2.87$, while at $R_{\rm s} \le 0.15$ the second bound
state appears in the near-barrier well.

Independently of \cite{konst-shik}, the idea of depletion region
near the barrier was also suggested in \cite{appel-baraff}, where
the energy of localized states and the depth of self-induced
potential well were calculated as functions of $R_{\rm s}$. However,
in these works either the self-consistency has not been reached, as
in \cite{konst-shik}, or some model potential was used in the
Schr\"{o}dinger equation instead of the exact solution of Poisson
equation \cite{appel-baraff}. The exchange-correlation interaction
of free charge carriers was neglected in both papers. The
self-induced levels are quite shallow with energies less than $0.1$
of the potential well depth ($\simeq 2$ meV for a semiconductor with
n-GaAs parameters), thus being very close to the states of
continuous spectrum. A small change in the potential well shape can
make them disappear. In this context, it was interesting to find out
if these states remain in the self-consistent solution.

The Kohn-Sham set of equations was solved according to iteration
scheme \eqref{N_qu-init} - \eqref{N_qu-next}. The total
dimensionless electron density $n^i$ at $z\geq d/2$ obtained after
solving the Schr\"{o}dinger equation on the $i$-th iteration was
calculated by formula
\begin{equation}\label{N_loc_plus_continue}
n^i(\zeta)%
=3\displaystyle \int_{0}^{\rm 1}dk(1-k^2)|\psi_{k}^{i}(\zeta)|^2
+\frac{3\pi}{2}\sum_{j}|\psi_{j}^i(\zeta)|^2(1-\varepsilon^i_{j}),
\end{equation}
where $\zeta = k_{\rm F}(z -d/2)$, $j$ is the number of a discrete
level, $\varepsilon_{j}$ are discrete energy levels, $\psi_{k}$ and
$\psi_{j}$ are the respective wave functions of continuous and
discrete parts of the spectrum. The first term
in(\ref{N_loc_plus_continue}) corresponds to the density of
continuous-spectrum electrons, and the second one is the electron
density in the localized states.

Since the considered problem with an infinitely high barrier is
close to that of potential scattering of $s$-wave by a short-range
potential, one could search for energy levels in the potential well
by means of the Levinson's theorem (see, for example,
\cite{Newton-66}) that expresses the number $m$ of levels in a well
in terms of the asymptotic phase \eqref{psi-asympt} of
continuous-spectrum wave functions at infinity by relation
$\lim_{k\rightarrow 0}\gamma_k=m\pi$. Although this formula is used,
for example, in \cite{baraff-appel} for a theoretical analysis of
barrier structure, it is not convenient in numerical calculations
because of redundant computation of wave functions for empty states.
Instead of this, it suffices to find a solution of the
Schr\"{o}dinger equation \eqref{schred^i} at $k=0$ and calculate the
number of zeros within the interval $0<\zeta <\infty$.

However, when finding the energy and the wave function of a bound
state we applied the trigonometric sweep method, which eliminates
the well-known numerical instability of solution decreasing in the
classically forbidden region, caused by the second solution that
grows exponentially \cite{Fedorenko-94}. In this method, the wave
function is written as
\begin{equation}\label{Eq psi_Asin}
    \psi(\zeta)=a(\zeta)\sin\eta(\zeta),\,\,\,\,
    \frac{d\psi}{d\zeta}=a(\zeta)\cos\eta(\zeta),
\end{equation}
so one can easily show that the total phase $\eta(\zeta)$ introduced
here has the increment $\pi$ at zeros of the function $\psi(\zeta)$
where the logarithmic derivative of the latter turns to infinity.
There are two first-order equations for the phase and the amplitude
\begin{align}
   & \frac{d\eta}{d\zeta}=\left(\varepsilon-u_{\rm eff}(\zeta)+u_{\rm    eff}(\infty)\right)\sin^2\eta + \cos^2\eta \label{Eq eta}\\
&\frac{d a}{d\zeta}=\frac{1}{2}a \sin 2\eta\left(1+u_{\rm
eff}(\zeta)-u_{\rm eff}(\infty)-\varepsilon\right), \label{Eq a}
\end{align}
which are derived from \eqref{schred^i} taking \eqref{Eq psi_Asin}
into account. The equation for $\eta(\zeta)$ is solved with initial
condition $\eta(0)=0$, while the initial condition for the amplitude
can be taken at random and specified definitely by the wave function
normalization to $1$. The level energy $\varepsilon_j \leq 0$
corresponds to the limit value of phase
$\eta_{\varepsilon}(\infty)=j\pi, (j=1,2,..)$. So to calculate the
number of levels in a well, one should only solve the equation
\eqref{Eq eta} for two energy values corresponding to the potential
well bottom and the edge of continuous spectrum $\varepsilon =0$,
and find the $\pi$ multiplicity of the difference between the two
phases at infinity.

\begin{figure}[h]\center
  \includegraphics[scale=0.6,  ] {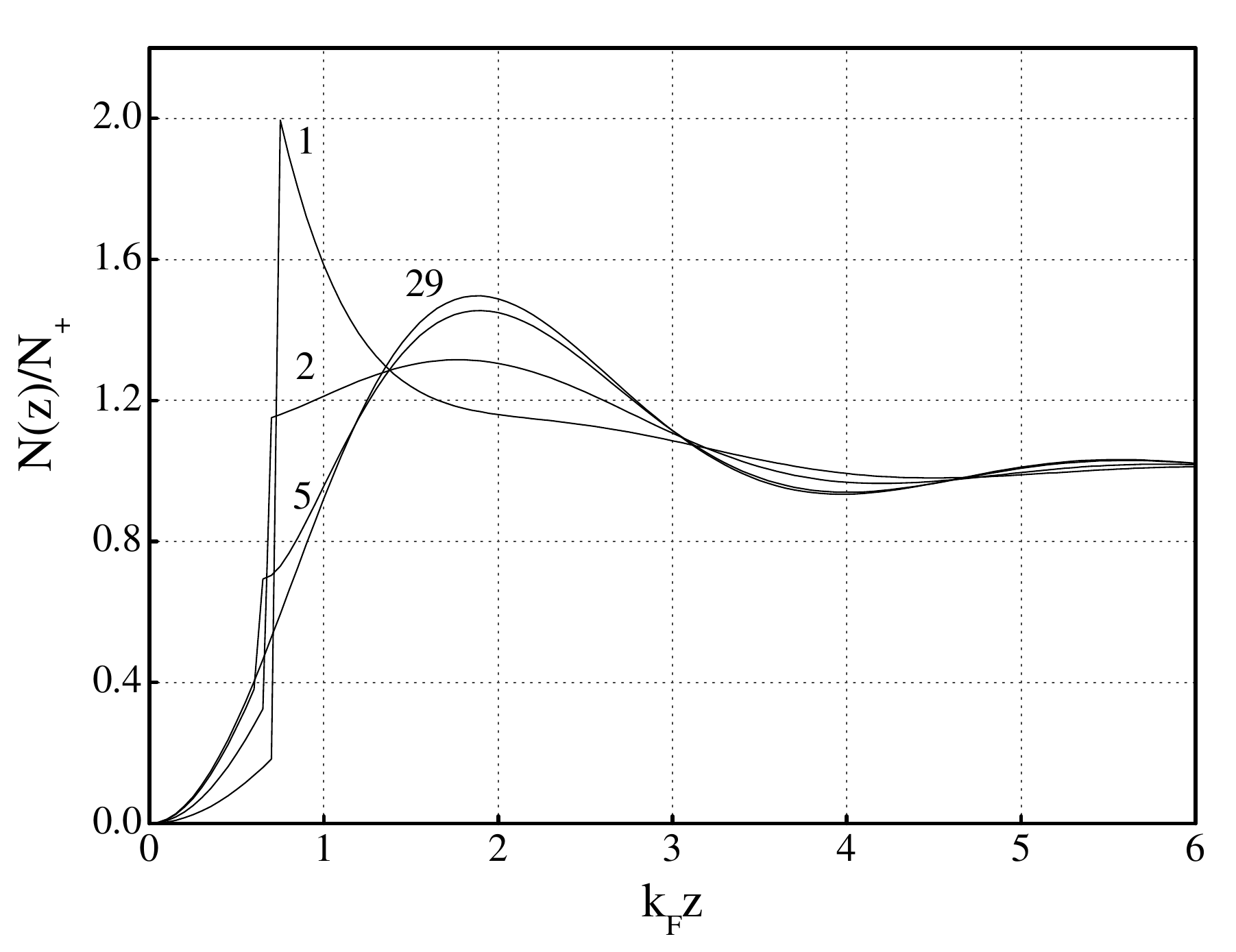}\\
  \caption{Electron density obtained at various cycles of iteration process. Number of
  curve corresponds to the number of iteration after which it was calculated. In this section, coordinate
  $z$ in the figures is counted from the point $d/2$.}
 \label{fig1-n(z)-jump}
\end{figure}

Boundary conditions for the Poisson equation were taken as
\begin{equation} \label{Eq_Boundary cond_levels}
\left.\frac{du}{d\zeta}\right|_{\zeta=0}=0,\qquad%
\left.\frac{du}{d\zeta}\right|_{\infty}=0.
\end{equation}
Note that such boundary conditions along with Fermi level specified
by the requirement of local neutrality at $\zeta \rightarrow \infty$
and the choice of $u(\infty)=0$ define a unique numerical solution
of the Poisson equation \eqref{Poisson-crit point} and ensure its
tending to zero at infinity.

Fig.\ref{fig1-n(z)-jump} demonstrates the electron density
distributions obtained by various number of iterations. One can
easily see that, at the inintial iterations, the spatial dependence
of electron density has discontinuities at the boundary of validity
of quasiclassical expression for induced charge in the form
(\ref{N_ind-def}). The discontinuities disappear in the course of
iteration process, and final self-consistent density is a smooth
function of coordinate. It is also essential to note that iteration
convergence criteria specified were usually reached at 10-14 cycles
(depending on $R_{s}$). The continued calculations did not provide a
noticeable increase in accuracy at chosen discretization parameters,
but neither disrupted the solution, as can be seen from Fig. 1,
which testifies to the stability of the algorithm applied.

\begin{figure}[h]\center
  \includegraphics[scale=0.4,   ]{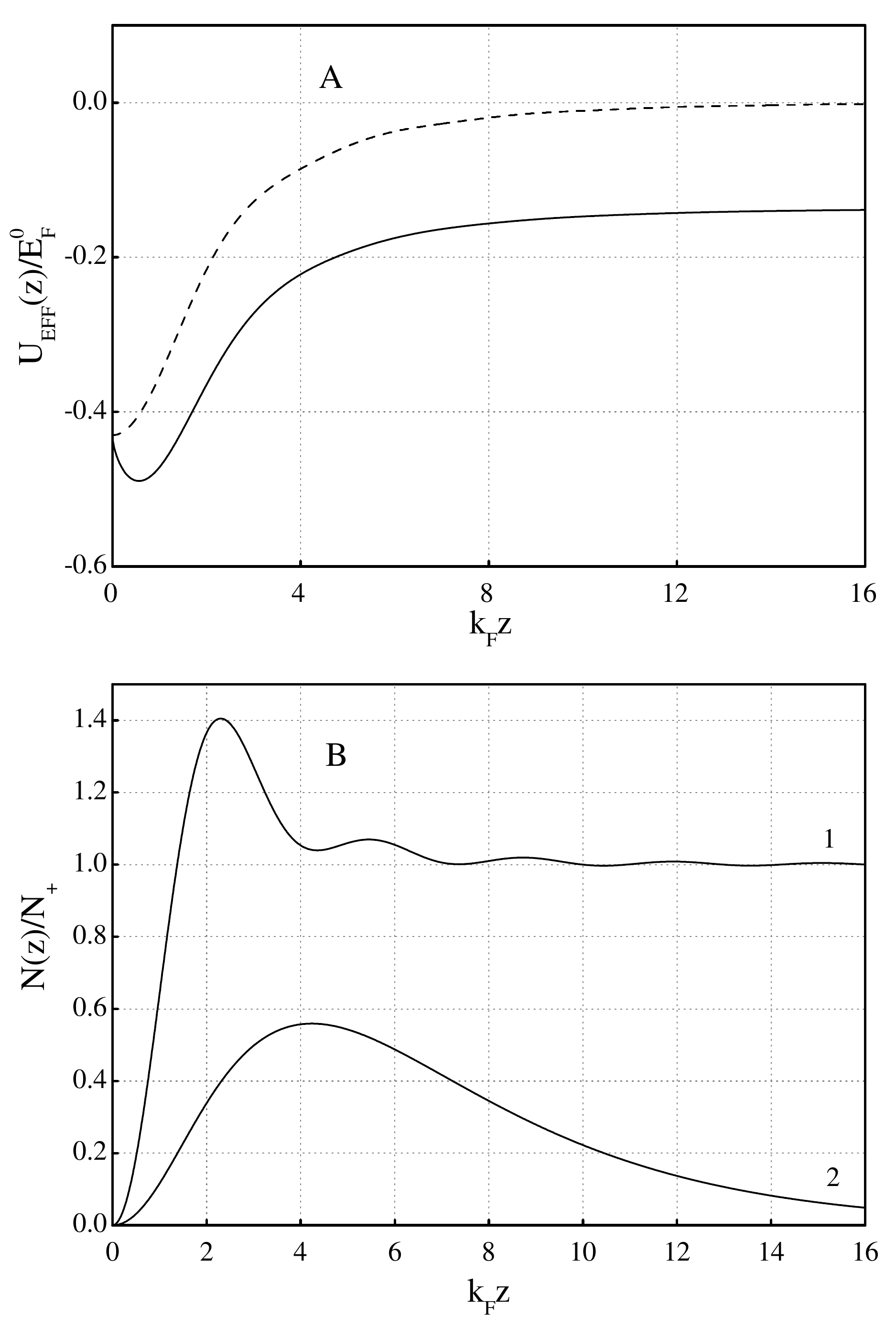}\\
  \caption{Self-consistent potential and electron density at $R_{s} = 0.4$.
  Fig. 2A demonstrates the self-consistent potential with account of the
  exchange-correlation interaction (solid line) and that
  in the Hartree approximation (dashed line). Fig. 2B shows the total
  self-consistent electron density (curve 1) and density of electrons
  localized in the size-quantized subband (curve 2).}
  \label{fig2_n-eff-potwell}
\end{figure}

\begin{figure}[h]\center
  \includegraphics[scale=0.5,   ]{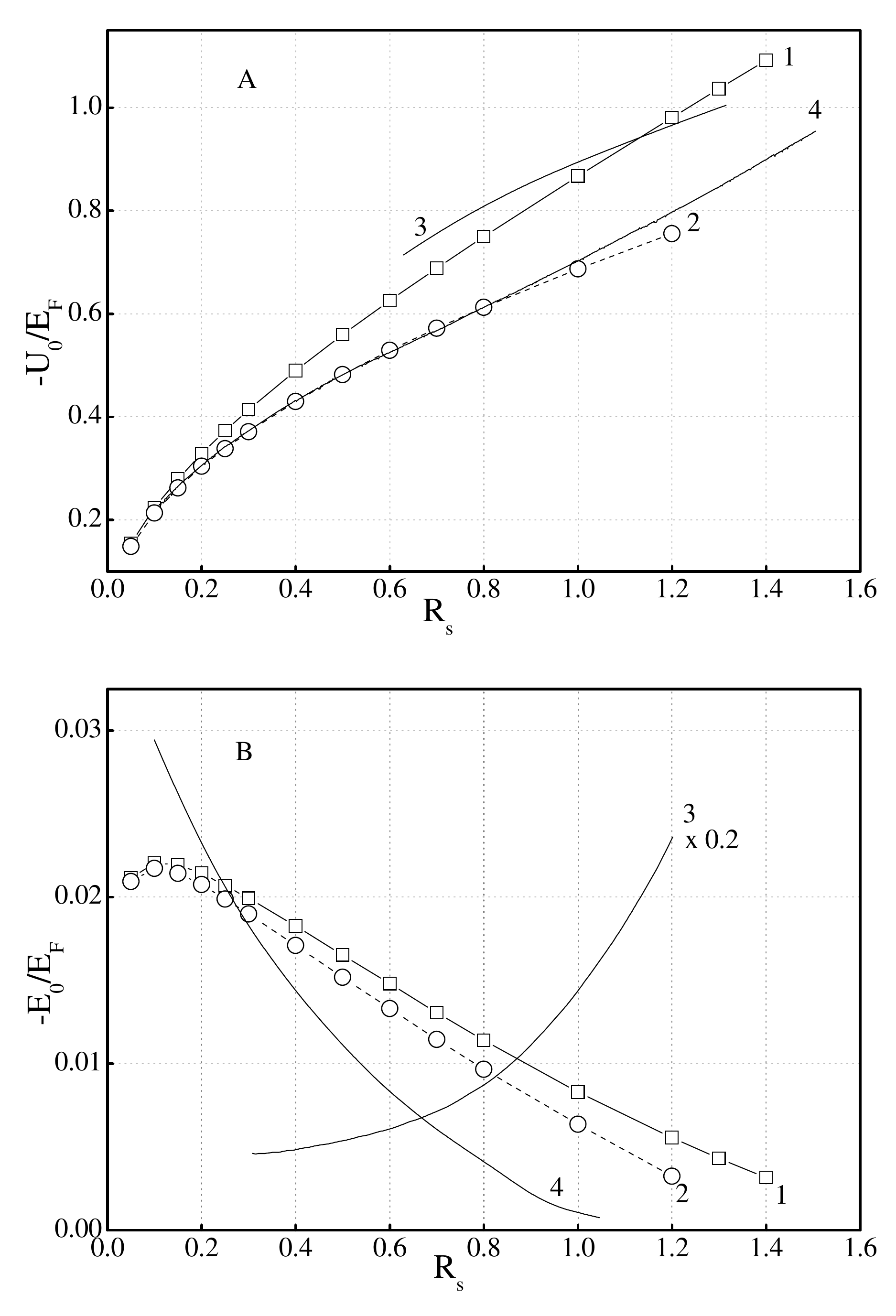}\\
  \caption{Potential well depth (A) and bound-state
  energy (B) as functions of $R_{s}$.  Curves 1 were calculated with account of
  the exchange-correlation interaction, curves 2 are results of
  self-consistent calculation in the Hartree approximation.
  Curves 3 and 4 represent results presented in \cite{konst-shik} and \cite{appel-baraff} respectively.}
  \label{fig3_compare}
\end{figure}

Fig. \ref{fig2_n-eff-potwell} demonstrates the coordinate dependence
of the self-consistent potential and electron density. As can be
seen from Fig. \ref{fig2_n-eff-potwell}A, taking into account the
$U_{\rm xc}$ modifies shape and depth of the potential well and,
correspondingly, the energy of localized state (see curves 1 and 2
in Fig. \ref{fig3_compare}). The total self-consistent electron
density in Fig. \ref{fig2_n-eff-potwell}B (curve 1) is the sum of
density of electrons in the size-quantized subband (curve 2) and
that of electrons in the continuous spectrum states. Localized
electrons are concentrated mainly close to the potential barrier,
though the localization pattern depends on $R_{\rm s}$.

The $R_{\rm s}$ dependencies of self-consistent well depth and
energy of the state localized therein are presented in Figs.
\ref{fig3_compare}A and \ref{fig3_compare}B respectively. Curves 1
and 2 in both figures are results obtained in present work, while
curves 3 and 4 are results presented in papers \cite{konst-shik} and
\cite{appel-baraff}. One can see that calculated level energy and
its $R_{\rm s}$ dependencies presented in \cite{konst-shik} (curve
3B) differ both quantitatively and qualitatively from our
self-consistent calculations. Furthermore, we did not found two
localized levels in the potential well in the absence of external
electric field at any values of $R_{\rm s}\geq 0.05$. The second
level with a dimensionless cohesive energy about $0.0016$ was found
only at $R_{\rm s}=0.005$ that is not, as a rule, realized
physically in real structures. Nevertheless, one should note that
the idea of a self-consistent potential well with bound states close
to infinitely high barrier, put forward in \cite{konst-shik}, proved
to be true.

The results reported in \cite{appel-baraff} for the potential well
depth are in rather good agreement with the self-consistent
calculation of the present work in the Hartree approximation (see.
curves 2 and 4 in Fig. \ref{fig3_compare}A). However, the energy of
bound state (curve 4 in Fig. \ref{fig3_compare}B), calculated in
\cite{appel-baraff} differs from the accurate solution both in
magnitude and the $R_{\rm s}$ dependence, particularly at small
values of this parameter. At the same time, the difference cannot be
attributable to the exchange-correlation interaction neglected in
\cite{appel-baraff}, as it is evident from the relative closeness of
our curves 1 and 2 in Fig. \ref{fig3_compare}B to each other.

It may be safely supposed that considerable error in the
self-consistent energy levels found by method used in
\cite{appel-baraff} is caused by inaccurate description of the
potential well shape by a three-parameter model potential.

Taking into account the finite barrier height leads to a change in
boundary condition for electron wave functions at the
insulator/conductor interface, and that, as calculation shows, leads
in its turn to a lowering upper limit of $R_{\rm s}$ region where
localized states are observed. At a considerable decrease of barrier
height the discrete level disappears entirely.

\begin{figure}[h] \center
  \includegraphics[scale=0.6,   ]{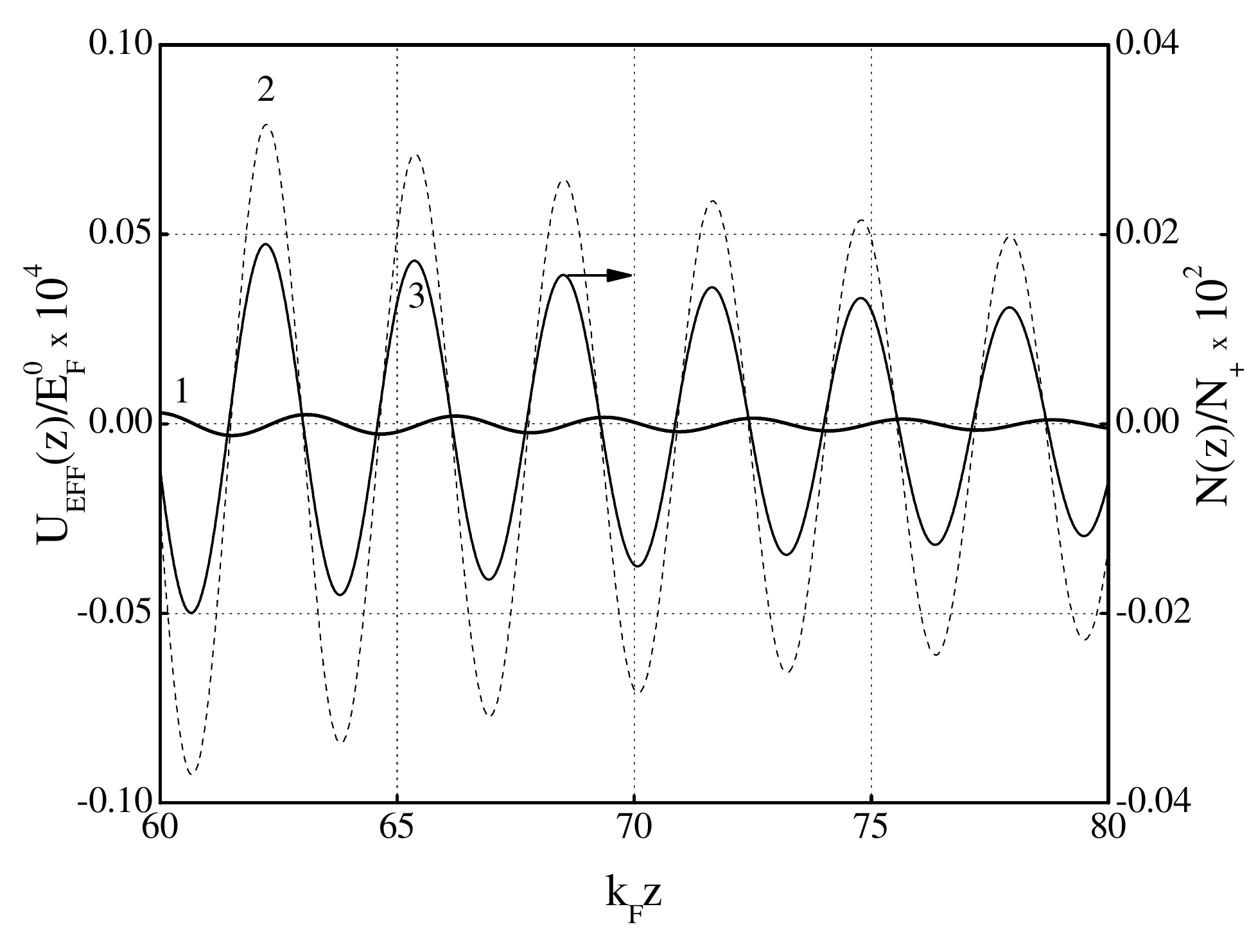}\\
  \caption{Friedel oscillations of effective potential and
  electron density in the electron gas limited by an infinitely high barrier.
  Curve 1 is effective potential, curve 2 is the Coulomb component of effective
  potential, curve 3 is electron density}
  \label{fig4_friedel}
\end{figure}

Far from potential barrier, $k_{\rm F}z \gg 1$, the effective
potential and electron density manifest the Friedel oscillations
(see. Fig. \ref{fig4_friedel}). One can see that the oscillation
amplitude of effective potential with $U_{\rm xc}$ component is
considerably less than that of Coulomb potential. Such suppression
of oscillations can be understood as follows. According to Poisson
equation, the oscillating component of electron density is maximal
at the same points as the Coulomb potential. The
exchange-correlation contribution into $U_{\rm eff}$ always has a
sign opposite to that of the direct interaction. So in virtue of the
repulsive nature of electron-electron interaction, the potential
$U_{\rm xc}$ is negative, and its magnitude grows steadily with
electron density according to formulae (\ref{Pot_xc-def}),
(\ref{U_x-def}) and (\ref{Uc-expli}). Consequently, at the points of
maximum Coulomb potential the exchange-correlation contribution is
also maximal in magnitude and negative.

The similar considerations show that at the minima of the effective
potential the Coulomb contribution and the exchange-correlation one
also summarize in phase opposition. As long as the relative role of
$U_{\rm xc}$ grows with decreasing electron density, this leads to
the oscillation magnitude of effective potential being suppressed
more than that of the Coulomb one with growing $R_{\rm s}$. Note
that the influence of $U_{\rm xc}$ upon the $R_{\rm s}$ dependence
of oscillation magnitude of the dimensionless electron density is
quite opposite. The reason for this is discussed in more detail in
Section \ref{Sec-Schottky}.

In conclusion, one should note that the system considered with
filled states of discrete and continuous spectrum has much in common
with quasi two-dimensional electron gas of accumulation layers close
to semiconductor-insulator boundary with an external electric field
applied or even without it. Though one believes \cite{AFS-RMP82}
that the problem of self-consistent spectrum calculation for
size-quantized subbands in the enriched layer is solved in
\cite{baraff-appel}, publications concerned with methods for
calculating such structures still appear \cite{Sune ea_JAP91}.
However, as before, no proved technique is suggested allowing to
avoid the mismatch between the total charge or the dipole moment of
quasi two-dimensional gas obtained at intermediate steps of
iteration and boundary conditions to Poisson equation imposed on the
potential. To make a solution possible, the authors renormalize the
electron density found by means of the Schr\"{o}dinger equation, and
also vary the Fermi level position in the bulk of semiconductor. The
results obtained in the present section of our work make clear that
the algorithm suggested allows one to obtain the solution in a
regular manner within the framework of mathematically complete and
accurate iteration procedure.

\subsection{The capacitance limitation of a barrier structure}
\label{SSec C-limit}

Up to now it was assumed that external voltage is not applied to
barrier structure. In this section, we find the self-consistent
distribution of electron density and effective potential in the case
of a dc bias $V$, and calculate the capacitance of the barrier
structure. Apart from the check for the algorithm performance, this
problem is of independent interest in connection with the
capacitance limitation of thin-film capacitors in general, posed in
\cite{Mead-PRL1961}, as well as with particular case of
nanocapacitor with a thin-film ferroelectric insulator in view of
its application prospects in electronics
\cite{Stengel-Spaldin_06_C-nano}.

It was shown by Mead \cite{Mead-PRL1961} that the measured
inverse capacitance of the metal-insulator-metal junction as a
function of insulation layer thickness $d$ does not tend to zero at
$d\rightarrow 0$ in contrast to the common formula for geometrical
capacitance of a plate capacitor. This was explained qualitatively
by the finite thickness of the spatial charge layer that screens the
electric field penetrating the real metal. The quantitative analysis
of this situation in the framework of the Thomas-Fermi approximation
for electrons in the metal was carried out in \cite{Ku-Ull_JAP64}
and complemented with an account of dielectric permeability of the
metal electrode in \cite{Black-Wels_ED99}. In the latter work, it
was also noted that using the insulator with a high permeability
$\kappa_{\rm d}$ makes the effect of non-ideality of electrons more
pronounced even at insulator thickness much greater than the
Thomas-Fermi screening length in the electrodes, and the maximum
achievable junction capacitance with given electrodes was introduced
as the limit at $\kappa_{\rm d}\rightarrow \infty$.

The  formulae for capacitance obtained in the above papers provided
a qualitative understanding of its experimental dependence on the
material of electrodes and insulator, and also allowed one to get a
correct order-of-magnitude estimate for the contribution of spatial
charge region to the capacitance. However it was noted that the
effect of the exchange-correlation interaction of electrons in metal
and that of the charge in the Friedel oscillations of the density
are still unclear and cannot be accounted for in terms of the
Thomas-Fermi approximation. At the same time, as early as in work
\cite{Bardeen-PhRev49} by Bardeen it was actually shown that spatial
electron redistribution related to the density oscillations caused
by the barrier does not keep the total neutrality of the system
because of the occurring deficit of electron charge. In the case of
an infinitely high barrier without an external field applied this
charge deficit equals (in the approximation of non-interacting
electrons)
\begin{equation}\label{Bardin charge}
\int\limits_0^\infty  {dz\left( {N(z) - N_+ } \right) =  -
\frac{{k_{\rm F}^2 }} {{8\pi }}},
\end{equation}
which coincides with the excess of the positive charge given by
formula (26) in \cite{Bardeen-PhRev49} (with printing error
corrected, see. Eq. (2) in \cite{Hunt-51PR}). Besides, it follows
from the results of previous section that self-induced surface level
also can capture a considerable electron charge (see. Fig.
\ref{fig2_n-eff-potwell}B). With a noticeable field dependence of
charge in the oscillations and on the near-barrier level their
contribution to the capacitance may be comparable with that of the
charge in the Thomas-Fermi screening region.

All above factors can be estimated by solving self-consistently the
Kohn-Sham equations for a barrier structure in the jellium model.
Let us consider a system of two semi-infinite metals separated by
the infinite potential barrier of width $d$ and dielectric
permeability $\kappa_{\rm d}$. Such a barrier prevents the charge
transfer between two parts of the system when voltage is applied.
The differential electrostatic capacitance per unit area of a wholly
neutral system is given in terms of the charge $Q$ per unit area of
the right-hand side of the structure by formula
\begin{equation} \label{Eq Capac-def}
C(V) =-
\frac{dQ}{dV}=\frac{d}{dV}\int_{z_0}^{\infty}dz(N(z,V)-N_{+}(z)),
\end{equation}
where $V=\varphi(-\infty)-\varphi(\infty)$ is voltage across the
structure equal to the work done by the electric field on a unit
positive charge to move it from negative to positive infinity. Here
$\varphi(z)$ is the dimensional Coulomb potential. The point $z_0$
given by $dD/dz|_{z_0}=0$ is the sign changing point of charge
density in the region where electric induction $D$ is maximal. As
long as electric field is zero at both infinities because of the
finiteness of potential $\varphi(z)$, the definition \eqref{Eq
Capac-def} provides that charges in two parts of the structure are
equal in absolute magnitude.

We assume that the metal electrodes are identical and the positive
background distribution is described by formula \eqref{Eq. barrier
N+}. To simplify calculation, we also neglect a possible nonzero
value of the contact potential difference along with related
built-in fields and charges in surface states, as it was assumed in
the model considered in \cite{Ku-Ull_JAP64} and
\cite{Black-Wels_ED99}. These conditions mean the absence of charge
within the barrier, which allows one to take any point of the
insulator as $z_0$ including the boundary ones $z_0=\pm d/2$. If the
induction $D$ inside the barrier is specified, the right-hand
electrode charge equals
\begin{equation}\label{Eq Q-right}
    Q(V)=-D/ 4\pi.
\end{equation}
We represent the voltage across the total structure as a sum of
voltages across its three constituent parts,
\begin{equation}\label{Eq V_L-R_def}
    V=[\varphi(-\infty)-\varphi(-d/2)]+ E_{\rm d} d
    +[\varphi(d/2)-\varphi(\infty)],
\end{equation}
where field inside the barrier $E_{\rm d}=D/\kappa_{\rm d}$. Let
$\varphi_{\pm}(z,E)$ be the Poisson equation solutions at $d/2 \leq
z \leq \infty$ and $-d/2 \geq z \geq -\infty$ with boundary
conditions $\-d\varphi_{\pm} / dz|_{z=\pm d/2} = E, \,\,\,
\varphi_{\pm}(\pm \infty)=0$ respectively. Then, in view of
induction continuity at the insulator-metal boundary, the equality
\eqref{Eq V_L-R_def}, can be rewritten as
\begin{equation} \label{Eq V_L-R_redef}
    V=-\varphi_{-}(-d/2,D/\kappa_{\rm m}) +E_{\rm d} d  + \varphi_{+}(d/2,D/\kappa_{\rm m}).
\end{equation}
Let $E_{\rm m}=D/\kappa_{\rm m}$ be the field in metal at the
insulator interface, where the not-equal-to-one dielectric
polarization $\kappa_{\rm m}$ takes into account a possible
polarizability of internal electron shells of ion cores. Since in
our case $\varphi_{-}(-d/2,E_{\rm m})=\varphi_{+}(d/2,-E_{\rm m})$,
we get a resulting formula, which reduces the calculation of
barrier-structure capacitance to the problem for semi-infinite
system considered in the previous section,
\begin{equation}\label{Eq V_L-R}
    V=\varphi_{+}(d/2,E_{\rm m})-\varphi_{+}(d/2,-E_{\rm m}) + E_{\rm d} d .
\end{equation}
We calculate the capacitance at small voltage when differentiation
in Eq. \eqref{Eq Capac-def} can be replaced with the ratio of small
increments. Due to the additive structure of $V$ in \eqref{Eq
V_L-R}, it is more convenient to deal with the inverse capacitance.
Taking \eqref{Eq Q-right} into account, we get
\begin{equation}\label{Eq C_tot}
    C^{-1}=-\frac{V}{Q}=4\pi\frac{\varphi_{+}(d/2,E_{\rm m})-\varphi_{+}(d/2,-E_{\rm m})}{\kappa_{\rm m}
    E_{\rm m}}+4\pi \frac{d}{\kappa_{\rm d}} \equiv 2C^{-1}_{\rm i}+ %
    C^{-1}_{\rm d},
\end{equation}
where the standard notation $C_{\rm i}$, $C_{\rm d}$ is used at the
end for separating contributions of voltage across the interfaces
(inside the metal electrodes) and the insulator to the total
capacitance. The explicit definition for $C_{\rm i}$ and $C_{\rm d}$
can be easily seen from the structure of expression \eqref{Eq
V_L-R_redef}, if one takes into account that $\varphi_{+}(d/2,0)\neq
0$ (see. Fig. \ref{fig2_n-eff-potwell}A). It is evident that the
maximal possible capacity $C_{\rm max}$ of the considered structure
is achieved in the limit of small electric thickness of insulator,
$\lim d/ \kappa_{\rm d} \rightarrow 0$, and can be written as
\begin{equation}\label{Eq C_max def}
C^{-1}_{\rm max}=
4\pi\frac{\varphi_{+}(d/2,E_{\rm m})-\varphi_{+}(d/2,-E_{\rm m})}%
{\kappa_{\rm m} E_{\rm m}}.
\end{equation}
We assume further on that $\kappa_{\rm m}=1$, for, as may be
required, the numerical value of capacitance being found can be
easily recalculated using Eq. \eqref{Eq C_max def} and taking into
account the relevant redefinition of atomic units.

The structure of formula \eqref{Eq C_max def} shows that, for
calculating the capacitance limitation of barrier structure one
should solve twice the problem for semi-infinite metal bounded by an
infinitely high barrier with boundary conditions as
\begin{equation} \label{C-barr_Boundary cond}
\left.\frac{du}{d\zeta}\right|_{\zeta=0}=\pm
E,\qquad\left.\frac{du}{d\zeta}\right|_{\infty}=0.
\end{equation}
Here we use the dimensionless variables for electron potential
energy and spatial coordinate introduced in Section
\ref{SSec-Algorithm/Applicat}. The dimensionless electric field $E$
is normalized to intrinsic field $E_{\rm c}=k_{\rm F}\varepsilon
^0_{\rm F}/e$. Apart from the change in the boundary condition at
zero compared to \eqref{Eq_Boundary cond_levels}, no modifications
were needed in the calculation program.

The found values of potential at the insulator boundary at two
values of electric field were used in formula \eqref{Eq C_max def}
for capacitance calculation. For the purpose of accuracy and to
verify that the chosen field magnitudes are sufficiently small and
do not exceed the limits of linear voltage dependence of the charge,
the potential was calculated for three pairs of values $\pm E$, and
the slope of the linear dependence obtained was used to calculate
the capacitance.

\begin{figure}[h] \center
  \includegraphics[scale=0.6,   ]{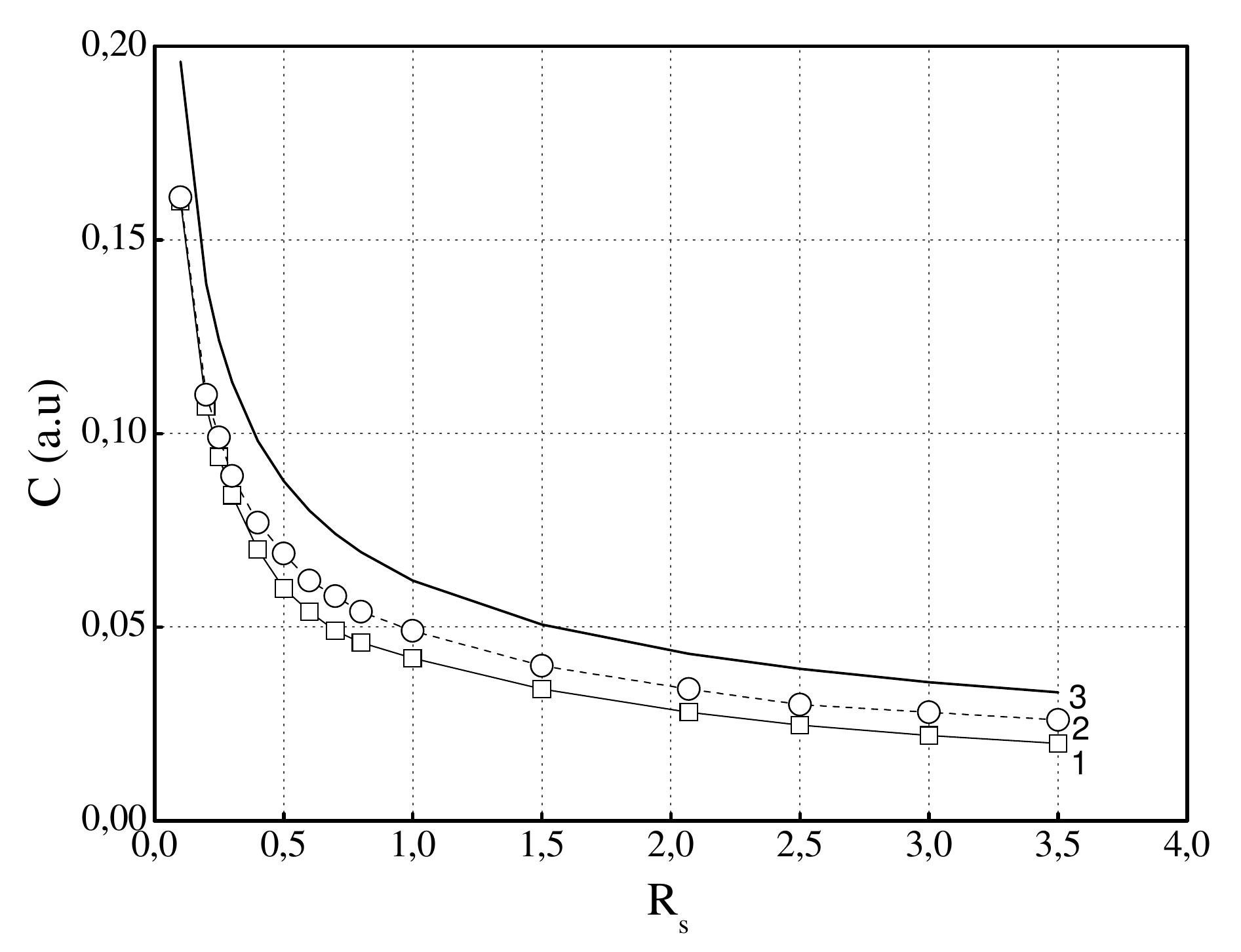}\\
  \caption{Capacitance of barrier structure at various values of $R_{s}$.
  Curve 1 is self-consistent calculation with account of exchange-correlation interaction,
  curves 2 and 3 are the same in the Hartree and Thomas-Fermi approximations respectively.}
  \label{fig_capac}
\end{figure}
Fig. \ref{fig_capac} demonstrates the $R_s$ dependence of
capacitance limitation (denoted just as $C$) of the structure
obtained in different approximations. Curves 1 and 2 correspond to
the self-consistent calculation of capacitance with account of the
exchange-correlation interaction and in the Hartree approximation.
Curve 3 was obtained in the Thomas-Fermi approximation and can be
represented by a simple explicit formula
\begin{equation}\label{C_T-F}
    C=\frac{k_{\rm TF}}{8\pi}=\frac{1}{8\pi}%
    \left(\frac{12}{\pi}\right)^{1/3}\frac{1}{R_{\rm s}^{1/2}},
\end{equation}
where $k_{\rm TF}$ is the inverse Thomas-Fermi length. If, in the
Thomas-Fermi approximation, we also take into account the
exchange-correlation potential in accordance with formula
\eqref{N_ind-def}, we get the capacitance formula like
\eqref{C_T-F}, but with $k_{\rm TF}$ replaced with renormalized
value $\tilde{k}_{\rm TF}=k_{\rm
TF}/(1+\frac{3}{2}du_{xc}/dn)^{1/2}$. The corresponding curve is not
shown in Fig. \ref{fig_capac} because the difference between these
two curves in the considered range of $R_{\rm s}$ is almost just as
small as between curves 1 and 2. With an exact calculation, the
account of exchange-correlation potential leads to the decrease of
capacitance, which is probably related to withdrawal of the minimum
of self-consistent potential well and charge accumulated therein
from the barrier (see Fig. 2a).

The comparison of numerical values of interface capacity $C_i=2C$
obtained from curve 1 with its experimental estimations (see, for
example, \cite{Sinnamon ea_APL01}) shows an agreement in the order
of magnitude. Thus, not taking into account the dielectric
permeability of electrode metal, which differs noticeably from unity
for the noble \cite{Eherenrich-PR62} and transition \cite{Choi
ea_PRB06} metals, the range of $C_i$ change, according to curve 1 in
Fig. \ref{fig_capac}, is from 275 to 85 fF/${\mu}$$^{2}$ at $0.5
\leq R_{\rm s} \leq 3.5$. In terms of the effective electric
thickness $d_{\rm eff}=(4\pi C_i)^{-1}$ we get the range from
$0.035$ to $0.11$ nm. The analysis of current theoretical and
experimental data made in \cite{Rabe_06_dead-layer} led to the
conclusion in favor of imperfect screening in metal contacts as the
main reason for the so-called "dead layer" observed experimentally.
The latter determines the maximum achievable capacitance in
structures with ferroelectric insulator for a specified choice of
metal for electrodes.

The general conclusion from the data presented in Fig.
\ref{fig_capac} is that the Thomas-Fermi approximation with its
exponential distribution of spatial charge does provide the
order-of-magnitude agreement with complete self-consistent
calculation as concerns the capacitance limitation. Neither the
exchange-correlation interaction of electrons, nor density
oscillations or an account of the self-induced potential well with a
discrete energy level therein prove essential. The level that
appeared at $R_{\rm s}< 1.6$ did not affect the smoothness of the
curve $C(R_{\rm s})$, in accordance with result reported in
\cite{Kohn-Maj_PR65}. The similar absence of drastic changes in the
experimental curve $C(V)$ at changing number of size-quantized
levels in the accumulation layer of $n-InAs$ was reported in
\cite{Tsui-8PRB73}.

In work \cite{baraff-appel}, the preservation of continuous
dependence of self-consistent well potential on the external
electric field regardless of the bound state appearance or
disappearance was explained by forming of an "orthogonal hole" in
the spatial distribution of electrons filling the states of
continuous spectrum . Our numerical results demonstrate that this
phenomenon is determined by appearance or disappearance of one more
zero in the continuous-spectrum wave functions with the appearance
or disappearance of a bound state in the potential well. As a
consequence, filling of a bound state is accompanied, in a sense, by
redistribution of total electron density among its two components in
formula \eqref{N_loc_plus_continue}, rather than contribution from
the level is added to the existing one from the continuous spectrum.
On the contrary, failing to account for (or neglecting) the
contribution from a bound state, even quite shallow, in the
iteration process leads to formation of a shallow and wide potential
well with more levels appearing at the next iteration cycle, so the
iteration process fails to converge.

\section {Work function and surface energy density of metal in the model with homogeneous background}
\label{Sec Work function & surf energy}

In the model with homogeneous positive background (jellium model),
the metal work function and surface energy density were calculated
in a good many papers. However, the existing data reveal sometimes
not only quantitative, but also qualitative difference. We believe
that the reason for such divergence is related to the complete
self-consistence being not reached in the works under discussion
when calculating the distribution of electron density and effective
potential. Among the detailed and comprehensive works with
calculation technique described one could mention publications
\cite{L-K70PRB4555}, \cite{L-K-workfun} and
\cite{Perdew-Wang-workfun}, our results we will compare with. Note
that the purpose of this part of our work is to obtain an actually
self-consistent solution for one of the classical problems of the
theory of inhomogeneous electron gas and assess the importance of
self-consistency by comparing the results with those obtained by
other methods for the same problem statement. We do not aim at
improving the jellium model or the density functional theory for
calculating the surface properties of metals.

The problem statement here is generally similar to calculation of
barrier structure properties considered in the previous section.
However, in the case of metal the surface barrier is formed entirely
by the self-consistent field and does not contain any specified seed
potential. Besides, in metals in ordinary conditions, the values of
$R_{\rm s}$ calculated for charge carriers with free electron mass
in the medium with dielectric permeability $\kappa_{\rm m}=1$ exceed
considerably $R_{\rm s}$ values typical for semiconductor structures
and lie in the range $ 2 < R_{\rm s} < 5.7$. All the above raised
the interest for a calculation intended to explore the behavior of
convergence to the self-consistent solution and assess the algorithm
stability at $R_{\rm s}$ values exceeding those considered in
Section \ref{Sec_barrier structures}.

\subsection{Work function of simple metals}
\label{subsec_W simple met}

\begin{figure}[h]\center
  \includegraphics[scale=0.6, ]{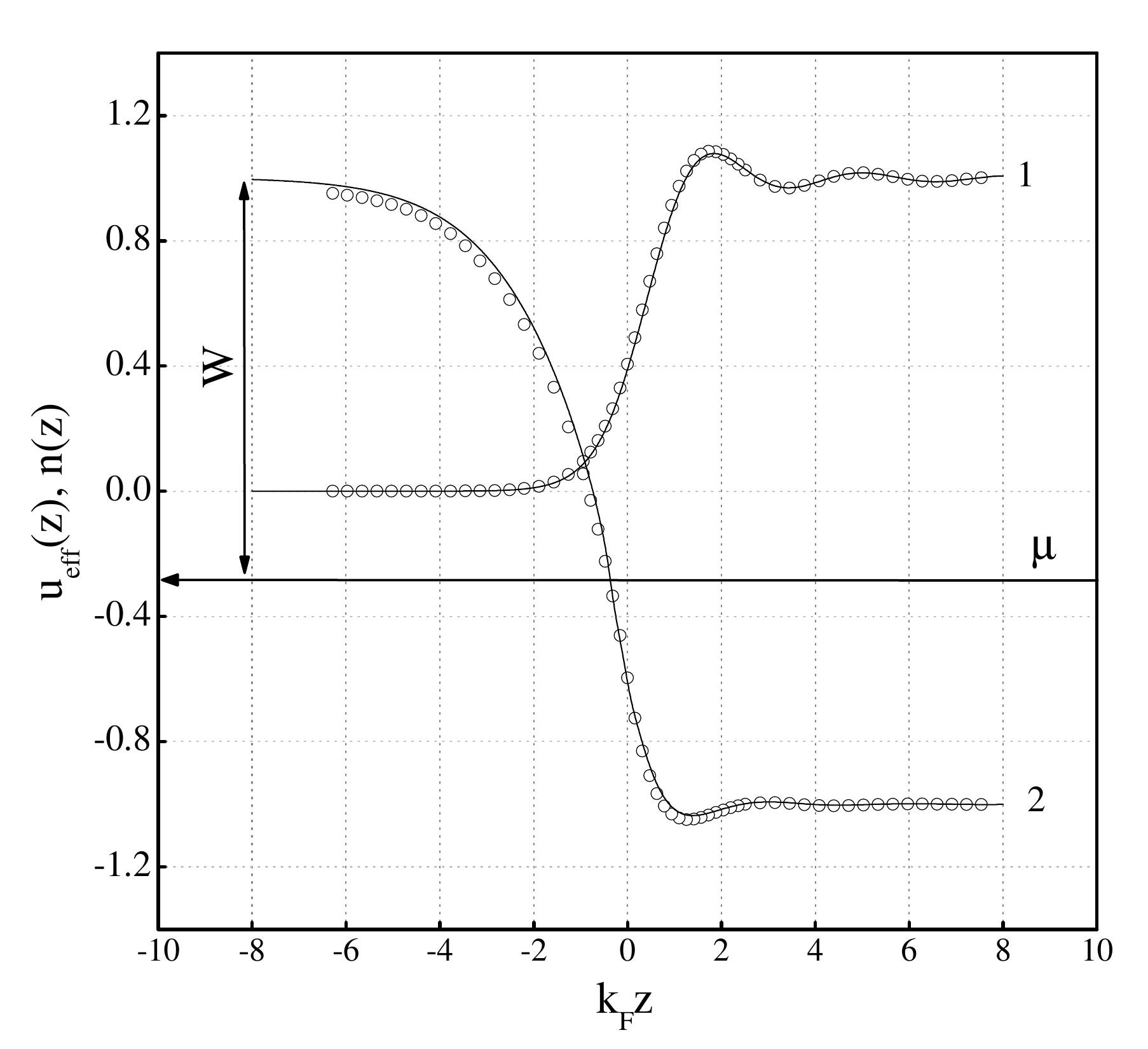}\\
  \caption{Comparison of self-consistent calculation for
  effective potential and electron density (solid curves) in present work with similar
  data from work \cite{L-K70PRB4555} (points) at $R_{s}=4$.}
  \label{fig_compare}
\end{figure}
Let the positive background occupy the half-space $z > z_{\rm +}$
where $z_{\rm +}$ is coordinate value at the metal-vacuum boundary.
In the calculation, the boundary conditions for the Poisson equation
\eqref{Poisson-crit point} were taken as $du/d\zeta|_{\zeta=\pm
\infty}=0$. Wave functions $\psi_k(\zeta)$ of single-particle states
with energies not exceeding the Fermi level were found as Cauchy
problem solutions exponentially damped at $\zeta\rightarrow -\infty$
and normalized according to Eqs. \eqref{full w-function} and
\eqref{psi-asympt}. The dimensionless work function
$w=W/\varepsilon_{\rm F}^0$ (see Fig. \ref{fig_compare}) equals the
potential barrier height counted from the value of bulk chemical
potential,
$$
w = u_{\rm eff}(-\infty) - \mu.
$$
The self-consistent electron density and effective potential
obtained after convergence of iteration process in accordance with
calculation scheme described in Section \ref{Sec-Algotithm} are
shown in Fig. \ref{fig_compare}. The curves $n$ and $u_{\rm eff}$
taken from the table 1 in the work \cite{L-K70PRB4555} are presented
by dots in the same figure for comparison. One should note that in
work \cite{L-K70PRB4555} the correlation energy was taken as
standard Wigner formula
\begin{equation} \label{Wigner_Ec}
\varepsilon_{c} = -\frac{0.44}{R_s + 7.8}\, ,
\end{equation}
rather than expression (\ref{E_c-def}) used in present work. For the
correct comparison, the result of our self-consistent solution
presented in Fig. \ref{fig_compare} was calculated using correlation
energy (\ref{Wigner_Ec}). On can see that calculation results in the
figure agree rather well, though in work \cite{L-K70PRB4555} a
12-parameter formula for electron distribution was used instead of
searching for self-consistency, and those parameters were found by
some variational procedure (see. Appendix B in \cite{L-K70PRB4555}).

However, the convergence of our iteration solution and its good
agreement with the generally accepted calculation results from the
work \cite{L-K70PRB4555} by Lang and Kohn at a particular value of
$R_{\rm s}$ are not in themselves sufficient for a full assessment
of the self-consistency degree achieved. In the work
\cite{budd-73PRL} within the model of homogeneous background, an
universal relation for electrostatic potential was obtained,
\begin{equation}\label{Badd}
U(z_{+})-U(\infty) = N\frac{d\varepsilon}{d N} \equiv
\tilde{\Delta}_{\rm BV},
\end{equation}
where $\varepsilon$ is total energy per one electron in homogeneous
electron gas. For further convenience, we introduce the
dimensionless parameter of self-consistency $\Delta_{\rm BV} =
\tilde{\Delta}_{\rm BV}/\varepsilon_{\rm F}^0$ and write it as
\begin{equation}\label{Badd by E_F}
\Delta_{\rm BV}=\frac{2}{5}+2\left( \frac{4}{9\pi}
\right)^{2/3}R_{\rm s}^2 \left[U_{\rm xc}-\varepsilon_{\rm xc}
\right],
\end{equation}
where the expression in square brackets is left in atomic units.
\begin{table}[h]
\centering
\caption{The $R_{\rm s}$ dependence of parameters $\Delta$ and $\Delta_{\rm BV}$ as well as
work function $W$ and surface energy density $\Sigma$.
The number of iterations needed to reach self-consistency is also presented.} \label{Table-metal data}
\begin{tabular}{|l|c|c|c|c|c|}
\hline
$R_s$ & $\Delta$ & $\Delta_{\rm BV}$  & $W$ \quad(eV) & $\Sigma\quad (\rm erg/{\rm cm^2})$  & $N_{\rm iter}$ \\
\hline
0.3     & 0.3751 & 0.3751& 3.12 & -      &12\\
0.5     & 0.358  & 0.358 & 3.21 & -      &8\\
1.0     &0.317   &0.317  & 3.40 & -      &6\\
1.3     &0.291   &0.291  & 3.51 & -      &6\\
1.5     &0.274   &0.274  & 3.60 & -      &7\\
1.65    &0.261   &0.261  &3.63  & -      &7\\
1.8     &0.248   &0.248  &3.64  & -      &7\\
2.07    & 0.224  & 0.224 &3.60  & -1340  &9\\
2.3     & 0.204  & 0.204 &3.53  & -583   &8\\
2.5     & 0.186  & 0.186 &3.48  & -321   &12\\
3.28    & 0.115  & 0.115 &3.12  & 106    &9\\
3.99    & 0.051  & 0.051 &2.87  & 71     &17\\
4.96    &-0.047  &-0.047 &2.50  & 65     &28\\

 \hline
\end{tabular}\\
\end{table}
The relationship \eqref{Badd by E_F} was used to verify if the
calculated potential is actually self-consistent with the electron
density. Good agreement between $\Delta = u(\zeta_{+})-u(\infty)$
and $\Delta_{\rm BV}$ proves that a genuine self-consistent solution
of the Kohn-Sham equations was found (see. \cite{budd-73PRL}, p.
1430). The Table \ref{Table-metal data} presents the values of
$\Delta$ and parameter $\Delta_{\rm BV}$ corresponding to the
results of self-consistent calculation for specific $R_s$ and
$\varepsilon_{\rm c}$ values in accordance with formula
\eqref{E_c-def}. One can see that $\Delta$ and $\Delta_{\rm BV}$
match for all $R_s \leq 3.99$ to an accuracy better than $0.001$.
Contrary to work \cite{Perdew-Wang-workfun} where the criterion
(\ref{Badd}) was used explicitly in the process of solution
building, this relationship is never taken into account in our
algorithm. So its satisfaction proves that $n$ and $u_{\rm eff}$
obtained are genuine self-consistent solutions of Kohn-Sham
equations for assumed approximation for the exchange-correlation
potential of electrons.

One should note the sensitivity of the self-consistency criterion to
the values of potential at the positive background frontier. The
degree of agreement between $\Delta$ and $\Delta_{\rm BV}$ presented
in Table 1 is achieved only if the potential $u(\zeta_{+})$ value
taken is interpolated to the discretization half-step $\delta \zeta$
towards vacuum region. In the calculation, $\delta \zeta$ was chosen
within the interval $0.05 \div 0.005$.

A similar self-consistency test for electron density and potential
found by Lang and Kohn \cite{L-K70PRB4555} was carried out in the
work \cite{budd-73PRL}. The data presented there in Table 1 show
that, even in the case of improved (unpublished) results of Lang,
the self-consistency criterion \eqref{Badd} is fulfilled to an
accuracy better than the third decimal place only for $R_s \leq 4$.
With growing $R_{\rm s}$ the agreement is steadily worsening and at
$R_s=6$ the deviation reaches $0.01$. This means that at large
values of $R_{\rm s}$ close to the existence limit of real metals,
the parametrization scheme chosen in work \cite{L-K70PRB4555} and
the method of calculating the parameters are not quite an adequate
replacement for an accurate solution of the Kohn-Sham set of
equations. This fact can be considered as an indication of serious
difference in the behavior of electron density and potential at the
$R_{\rm s}$ values close to critical $R_{\rm sc}$, from the jellium
model solutions at higher densities of electron gas, which was
discussed in some measure in Section \ref{Sec-Algotithm}.

It follows from the Table 1 that in our calculations at $R_{\rm
s}\simeq 5$ the self-consistency criterion is also fulfilled worse
than at small values. In the region $R_{s}\leq 3.99$, the iteration
process was programmed to terminate if the discrepancy defined by
formula \eqref{Nevyazka_N-gen} becames less than specified value
$\delta = 10^{-5}$. However, at $R_{s}=4.96$ the algorithm of
critical point $\zeta_{c}$ autodetection described in Section
\ref{SSec-Algorithm/Applicat} failed to provide convergence of the
Poisson equation solution due to the steady movement of critical
point $\zeta_{\rm c}$ from the metal surface into the bulk. To
obtain a definite result in this case, one had to turn off, after a
number of initial iteration cycles, the program module that
redefined $\zeta_{c}$. Nevertheless, after such compulsory choice of
critical point, the electron density jump in its vicinity did not
drop lower than $3 \cdot 10^{-4}$ in the course of iterations. At
$R_{\rm s} > 5$ the convergence of iteration solution was not
achieved at all. It appears that the crucial factor here, as
concerns the algorithm operation, is that at $R_{\rm s} \rightarrow
R_{\rm sc}$ the effective potential \eqref{Ueff-def} turns out
almost equal to the local exchange-correlation one, and the
contribution of the long-range Coulomb potential to forming the
surface barrier becomes actually inessential.

\begin{figure}[h]\center
  \includegraphics[scale=0.6, ]{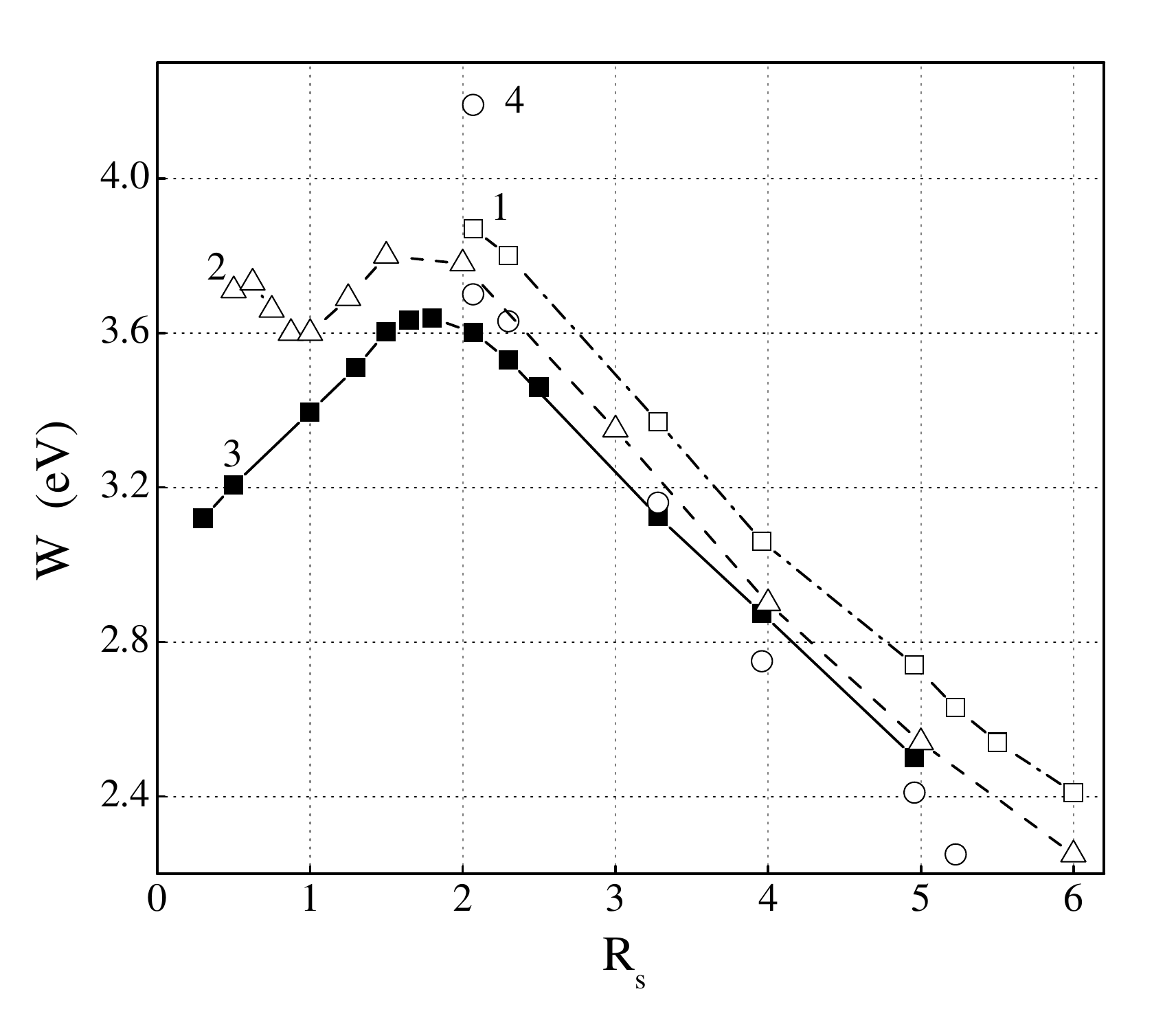}\\
  \caption{Work function in the homogeneous background model. Curves 1 and 2 represent  results of calculations
  in \cite{L-K-workfun} and \cite{Perdew-Wang-workfun}. Curve 3 was calculated in the present work.
  Circles 4 show experimental values of work function for polycrystalline specimens of
  Al, Li, Na, K and Rb from Table 3 of review \cite{Lang-83}. The smaller of two $W$ values
  at $R_s=2.07$ (Al) was measured in polycrystalline Al in the work \cite{Al-low WF_68FTT}.}
  \label{fig_workfun}
\end{figure}

The Table 1 demonstrates also the calculated $R_s$ dependence of the
work function $W$. In Fig. \ref{fig_workfun} this dependence (curve
3) is compared to similar ones presented in works \cite{L-K-workfun}
and \cite{Perdew-Wang-workfun}.
The values of work function calculated in present work are in rather
sensible agreement with experimental data for alkaline metals (Li
($R_s=3.28$), Na ($R_s=3.99$), K ($R_s=4.96$)) and lie closer to
measured values than those of curve 1 obtained in
\cite{L-K-workfun}. Most likely, the improved quantitative agreement
between our calculations and experiment is a consequence of using
the modified Wigner formula (\ref{E_c-def}) for correlation energy,
because at the same choice of expression for $\varepsilon_{\rm c}$
the results for the calculated potential are close to each other at
least for $R_s=4$, as can be seen in Fig. \ref{fig_compare}.

Our results for the work function at $R_s > 1$ are in
semi-quantitative agreement with those obtained in
\cite{Perdew-Wang-workfun} (Fig. \ref{fig_workfun}, curve 2), and
this agreement improves with growing $R_s$. One can see from the
figure that both curves have maxima at $R_s \approx 2$, however, at
small $R_s$ there is a qualitative difference. According to
\cite{Perdew-Wang-workfun}, work function has a local minimum at
$R_s \approx 1$, which is not confirmed by our calculation. This
local minimum in work function is possibly related to the absence of
self-consistency in the solution of Kohn-Sham equations that has
been obtained in work \cite{Perdew-Wang-workfun} for small values of
$R_s$.

The method of solution used in \cite{Perdew-Wang-workfun} and
described in detail in work \cite{Monnier-Perdew_78PRB} replaces the
Poisson equation with an integral equation for potential in some
finite vicinity of metal surface and applies boundary conditions
that do not fix the total charge (see. \cite{Monnier-Perdew_78PRB},
Appendix A). Along with the problems noted in Section
\ref{Sec-Intro} in relation to the replacement of Poisson equation
with integral one, such boundary conditions cannot provide a
self-sustaining consistency of the final result of iteration
procedure, as it was supposed in \cite{Perdew-Wang-workfun} (see the
text after formula (7)). As mentioned in \cite{Perdew-Wang-workfun},
the process failed to converge at all at $R_s < 0.5$. The reason for
this lies probably in the increasing role of Coulomb potential
compared to that of the exchange-correlation one and, therefore, the
increasing sensitivity to self-consistency with a rise in electron
density.

We note once more that the work \cite{Perdew-Wang-workfun} and ours
are identical in the physical problem statement and in the
formulation of initial equations except for different correlation
energy formula used. At the same time, we faced no crucial
difficulties with convergence at the region of small $R_s$ and were
limited only by growing requirements for computer memory and
increased computation time caused by widening vacuum region with a
non-zero electron density. As concerns the calculation problems in
the region of large $R_s$, they require a further analysis and
possibly relate in part to the use of local density approximation
for the exchange-correlation potential. On the one hand, a ground
for such premise can be found in the discussion on the application
limits of local approximation for exchange-correlation interaction
of electrons (see, for example, \cite{Jones-Gunnars_61RMP89} and
\cite{Kohn_71RMP99}). On the other hand, an attempt to apply
directly an iteration procedure to the Kohn-Sham equation with
potential $U_{\rm xc}$ in the absence of the Coulomb contribution to
the effective potential has revealed that iterations failed to
converge.

\subsection{Surface energy} \label{Ssec_Me-surf energy}
The self-consistent distributions of electron density and potential
along with single-particle wave-functions obtained in our work allow
to calculate the surface energy as a function of $R_{s}$. However,
the formulae for calculating surface energy available in the
literature use the finiteness of the region occupied by metal. On
the one hand, this allows to find the surface energy as a difference
between total energies of two states of the specimen before and
after dividing it into two parts. On the other hand, this involves
the quasidiscrete spectrum of eigenvalues and the necessity of
precise  evaluation of asymptotic dependence of total energy
components on the specimen size as it tends to infinity (see, for
example, \cite{Paasch-Hiet83}). To calculate the properties of an
infinite specimen, one should have expressions for the densities of
the corresponding components of total energy. In the standard
formula of the density functional theory for the total energy of a
finite specimen,
\begin{equation}
E_{\mathrm{tot}}\left(  N(\mathbf{r})\right) =T_{s}\left( N(\mathbf{r}%
)\right) +\int d\mathbf{r\varepsilon}_{\mathrm{xc}}[N({\mathbf r})]N(\mathbf{r})+\frac{1}%
{2}\int d\mathbf{r}d\mathbf{r}^{\prime}\frac{\left(  N(\mathbf{r}%
)-N_{+}\right)  \left(  N(\mathbf{r}^{\prime})-N_{+}\right) }{\left|
\mathbf{r-r}^{\prime}\right|},
\label{e_total}%
\end{equation}
where integration is implied over the volume, the second and third
terms describing the exchange-correlation and electrostatic energy
of a many-electron system already include density definitions of the
respective components. As concerns the total kinetic energy $T_{s}$
of non-interacting electron gas, it does not possess such a
property, being generally calculated by means of the formula
obtained in \cite{Hunt-51PR} that expresses the sum of
single-particle energies till the Fermi energy as the integral over
the spectrum of the phase of continuous-spectrum wave functions with
asymptotic form \eqref{psi-asympt}. Following the formula (50) in
\cite{K-V83}, we define the kinetic energy density at zero
temperature as
\begin{equation}\label{Eq. Tkin-density}
   t_{\rm s}(\mathbf r) = 2\!\!\int\limits_{\varepsilon \leq
\varepsilon _{\mathrm{F}}}\!\!\mathfrak{D}\left\{ \varepsilon
\right\} \varepsilon \left| \Psi _{\varepsilon }(\mathbf r)\right|
^{2} - U_{\rm eff}(\mathbf r)N({\mathbf r}).
\end{equation}
Using this expression, one can define the total energy density
\begin{equation}\label{Eq. Etot-dens}
   \epsilon_{\rm tot}(\mathbf r)=t_{\rm s}({\mathbf r})+\varepsilon_{\rm %
   xc}[N({\mathbf r})]N(\mathbf r)+\frac{1}{2}U(\mathbf r)\left(N(\mathbf r) -
   N_{+}(\mathbf r)\right)
\end{equation}
and, accordingly, obtain the expression for the total energy in the
volume $V$ selected out of the infinite specimen,
\begin{equation}\label{Eq. Etot-V}
    E_{\mathrm{tot}}(V)=\int\limits_{V}d\mathbf{r}\epsilon_{\mathrm{tot}%
}(\mathbf{r}),
\end{equation}
which can be applied for calculating the surface energy.

The surface energy \cite{L-K70PRB4555} in the jellium model becomes
negative at $R_{s}<2.5$, which means that metals with a high density
of conduction electrons ($2<R_{s}<2.5$) could not exist in such
model. The surface energy values obtained in \cite{L-K70PRB4555}
differ the most drastically from the measured ones for multivalent
metals, though even the case of monovalent metals of the first group
demonstrates the increasing worsening of agreement between
calculations and experiment with decreasing $R_{s}$. As it was
shown, the main reason is that the jellium model neglects the
discrete structure of the positive background, which is actually a
crystal lattice of ion cores (see, for example, discussion in
\cite{Lang-83}). Nevertheless, it was interesting to find out if the
use of self-consistent solution affects the $R_{s}$ dependence of
surface energy.

The surface energy was calculated using two expressions. One of them
is based upon the strictly thermodynamical definition of surface
energy as the component of total energy of a bulk homogeneous
system, which depends on its surface area $S$ rather than the volume
$V$. Using, for instance, the definition (3.82) from
\cite{Leont_T-ka83}, one can write
\begin{equation}\label{Eq. Es+Ev}
    E_{\rm tot}=S\Sigma+V\epsilon_{\rm tot},
\end{equation}
where $\Sigma$ is the surface energy density. In our case of
semi-infinite specimen, the volume of the considered part of the
system can be modified by receding from the surface to the bulk. Let
us define the energy per unit area of the system region up to
coordinate $z$ by formula
\begin{equation}\label{Eq. Etot(z)}
  E_{\rm tot}(z)=\int\limits_{-\infty}^{z}dz\epsilon_{\rm tot}(z).
\end{equation}
Since the energy density along with all variables defining it
becomes constant in the bulk, the value of $E_{\rm tot}(z)$ should
become a linear function of coordinate at sufficiently large $z$.
Having continued this linear dependence back to the boundary of
positive background, we obtain a finite value independent on the
volume, which provides the surface energy by definition. The above
can be expressed by simple formula
\begin{equation}\label{Eq. Sigma-lin}
   \Sigma=\lim_{z->\infty} \left[ E_{\rm tot}(z)-\epsilon_{\rm tot}(\infty)z \right].
\end{equation}
It is assumed here that the positive background boundary is at
$z=0$.

Such definition of surface energy is rigorous and, as a proper
characteristic of a thermodynamical system, independent of the
surface-forming process, which is sometimes modeled to determine the
surface energy. However, it requires an enhanced accuracy in solving
differential equations and calculating integrals, because the
formula includes a difference of two large values. This difficulty
is eliminated, if Eq. \eqref{Eq. Sigma-lin} is rewritten in the form
\begin{equation}\label{Eq. Sigma-integral}
    \Sigma=\int\limits_{-\infty}^{0}dz\epsilon_{\rm tot}(z)+%
     \int\limits_{0}^{\infty}dz\left(\epsilon_{\rm tot}(z)-%
     \epsilon_{\rm tot}(\infty)\right).
\end{equation}
With such a formula for the surface energy, the result proves to be
less sensitive to the accuracy of calculation, because the
integrands tend to zero when receding from the surface.

In the literature, another definition of the surface energy is
generally used, which represents it as a difference between the
total energy of the semi-infinite specimen and that of a half of the
infinite one (see, for example, \cite{Lang-83}). Let us represent
for convenience, as in work \cite{L-K70PRB4555}, the total surface
energy density as a sum of kinetic, electrostatic and
exchange-correlation components:
\begin{equation}\label{Eq. Sigma-3 parts}
    \Sigma = \Sigma_{\rm s} + \Sigma_{\rm es} + \Sigma_{\rm xc}
\end{equation}
In the dimensionless form and after integrating over the wave vector
parallel to the surface in formula \eqref{Eq. Tkin-density}, the
components of the surface energy density are written as

\begin{multline}
\sigma_{\rm s}=\frac{1}{2\pi^{2}}\int_{-\infty}^{\infty}%
d\zeta\int_{0}^{1} d k (1-k^{4}) \left(  \left| \psi_{k}(\zeta) \right|^{2}-\frac{1}%
{2}\theta(\zeta) \right)  -\label{Eq. sigma-s}\\
-\frac{1}{3\pi^{2}}\int_{-\infty}^{\infty}d\zeta \left\{ \left(
u(\zeta)+ u_{\rm xc} \left[ n(\zeta) \right] \right)
n(\zeta)-u_{\rm xc} \left[1 \right] \theta(\zeta) \right\} 
\end{multline}%
\begin{equation}
\sigma_{\rm xc}=\frac{1}{3\pi^{2}}\int_{-\infty}^{\infty}d\zeta
\left\{\varepsilon_{\rm xc}\left[ n(\zeta)\right]n(\zeta)
-\varepsilon_{\rm xc}\left[1 \right]\theta(\zeta)\right\}
\label{Eq sigma XC}%
\end{equation}
\begin{equation}\label{Eq sigma-es}
    \sigma_{\rm es} =\frac{1}{6\pi^2}\int_{-\infty}^{\infty} d\zeta
    \{n(\zeta)-1\cdot\theta(\zeta)\}u(\zeta).
\end{equation}

The comparison of Eqs. (4.13)-(4.15) with the formulae (45a)-(45c)
in work [34] reveals that only the expressions for the kinetic
energy contribution $\sigma_{s}$ are different. Using Eqs. (4.5) and
(4.11) helped to avoid additional complications by eliminating
integration of phase (2.37) of wave functions (2.35) over energy.

The results of surface energy calculation are presented in the Table
\ref{Table-metal data}. One can see that negative values of the
surface energy appear at the values of $R_s$ even slightly larger
than those reported in work \cite{L-K70PRB4555}. The main reason for
this is a smaller value of correlation energy given by formula
\eqref{E_c-def}. The general conclusion that the jellium model is
insufficient to describe the surface energy in the domain of
existence of real metals is still true.

\section {Calculation of self-consistent potential\\
of a Schottky barrier structure} \label{Sec-Schottky}

Up to now, we considered the semi-infinite systems with purely real
wave functions. As a consequence, such systems do not carry electric
current. In this section, a structure is analyzed whose
single-particle wave functions are limited and show oscillating
behavior in both infinitely far regions. Such boundary conditions
for the Schr\"{o}dinger equation solutions admit the complex
eigenfunctions and system states with the nonzero stationary
current. To avoid too laborious calculations in actually a model
situation, we take a structure such as a metal-semiconductor contact
with Schottky barrier. Because of a much larger free carrier density
in the metal the potential barrier is almost totally formed in the
semiconductor. This allows one to reduce the calculation part of the
problem to finding the wave functions and the potential in a
semi-infinite space and apply to this structure without any
essential modification the iteration algorithm described in Section
\ref{SSec-Algorithm/Applicat} for searching the self-consistent
electron density and the effective potential barrier.

We restrict ourselves to the equilibrium situation when no bias
voltage is applied to the structure and the current is zero. Let the
metal occupy the region with $z <0 $, the degenerate n-type
semiconductor with the ionized donor concentration $N_{\rm D}$
located at $z>0$. For definiteness, we assume that high density of
states at the metal-semiconductor boundary fixes the position of the
conduction-band bottom at the interface at energy $\Phi_{\rm s}$
reckoned from the Fermi level. Since the under-barrier electron
density in semiconductor near the interface is small, the effective
potential coincides with the Coulomb one, thus defining the first
boundary condition for electrostatic potential. The second boundary
condition for solving the Poisson equation in the semiconductor
region is specified by the requirement for the potential to be
bounded at infinity, which  means at the same time that the electric
field turns to zero therein. Reckoning the energy from the value of
potential in the bulk, we get the boundary conditions for solving
the Poisson equation in the semiconductor,
\begin {equation}
U(0) \equiv U_{\rm 0} = \Phi_{\rm s}+ \varepsilon_{\rm F}, \qquad
\left.\frac{d U}{dz} \right|_{z=\infty}=0
\end {equation}
Following the common practice, we neglect the penetration of
electric field into the metal. Thus, the electrostatic potential in
the metal is assumed to be constant and equal to its value at the
interface.

To describe the single-particle states of continuous spectrum in
such one-dimensional inhomogeneous structure, we use the results of
the analysis carried out in \cite{ASH06} of the Hilbert space
related to the Hamiltonian of infinite system. We take as a basis
the wave functions of the right-hand $\Psi^{\rm R}$ and the
left-hand $\Psi^{\rm L}$ states
\begin{align}
    \Psi^{\rm R}_{k,\mathbf k_{\|}}\left(z,\mathbf r_{\|}\right)=C_{k}^{\rm R}%
    \psi^{\rm R}_{k}(z)\exp\left({\rm i}\mathbf k_{\|}\mathbf r_{\|}\right)/2\pi,
    \label{Eq Psi_R-def}\\
    \Psi^{\rm L}_{q,\mathbf k_{\|}}\left(z,\mathbf r_{\|}\right)=
    C_{q}^{\rm L}\psi^{\rm L}_{q}(z)\exp\left({\rm i}\mathbf k_{\|}
    \mathbf r_{\|}\right)/2\pi,
\label{Eq Psi_L-def}
\end{align}
which are uniquely defined by their asymptotic behavior at infinity,
\begin{equation}
\label{Eq psi_R-def}
    \psi^{\rm R}_{k}(z) = \left\{
\begin{array}{llr}
e^{-{\rm i} kz} + r^{\rm R}_{k}e^{{\rm i} kz}& z \to & \infty\\
t^{\rm R}_k e^{-{\rm i} qz} & z \to & -\infty\\
\end{array}
\right.
\end{equation}
and
\begin{equation}
\label{psi_L-def}
    \psi^{\rm L}_{q}(z) = \left\{
\begin{array}{llr}
e^{{\rm i} qz} + r^{\rm L}_{q}e^{-{\rm i} qz}& z \to & -\infty\\
t^{\rm L}_q e^{{\rm i} kz} & z \to & \infty\\
\end{array}
\right.
\end{equation}

Using two quantum numbers $k$ and $q$ to describe the asymptotic
behavior of eigenfunctions in the right-hand and left-hand infinite
regions is related to the asymmetry of the considered barrier
structure even in the absence of bias voltage. This asymmetry is
caused by different positions of conduction-band bottom in the metal
and in the semiconductor with respect to the Fermi level. Besides,
even a symmetric structure becomes asymmetric when biased, so the
different quantum numbers are rather a rule than an exception. The
relation between $k$ and $q$ is found from the condition that both
characterize the eigenfunction pertaining to the same eigenvalue
$\varepsilon$ of the Hamiltonian,
\begin{equation}\label{Eq k-q def}
  \varepsilon-U_{xc}(\infty)=\frac{q^2+{\bf k_{\parallel}}^2}{2m_{\rm M}} + U_{\rm M} =
\frac{k^2 + {\bf k_{\|}}^2}{2m_{\rm S}},
\end{equation}
Here $U_{\rm M}$ is the position of conduction-band bottom in the
metal reckoned from that in the semiconductor, and indices M and S
denote the effective masses in metal and in semiconductor,
respectively. The right-hand equality in Eq. \eqref{Eq k-q def}
defines $q$ as an implicit function of $k$ (and vice versa).
Normalization constants $C_{q}^{\rm L}$ and $C_{k}^{\rm R}$ equal
$1/\sqrt{2\pi}$ if $\psi^{\rm L}_{q}$ are normalized to
$\delta(q-q')$ and $\psi^{\rm R}_{k}$ are normalized to
$\delta(k-k')$. The proof of mutual orthogonality of the pair of
functions $\Psi_{\varepsilon}^{\rm R,L}$ at one energy value
$\varepsilon$, which is necessary to have a full orthonormalized
basis in the Hilbert space of the problem, was presented in
\cite{ASH06}.

However, in the considered structure, all the results can be
expressed in values related to the semiconductor electrode of the
metal-semiconductor contact, so it is more convenient to normalize
$\psi^{\rm L}_{q}$ also to $\delta(k-k')$. In this case, for the
normalization constant of the left-hand states one should use the
expression
\begin{equation}
\left|  C_{k}^{\rm L}\right|  ^{2}= \frac{1}{2\pi}
 \frac{\partial q}{\partial k}.
\label{Eq C4L-k norm}%
\end{equation}

Solving the problems where the self-consistency is unnecessary, one
usually normalizes the wave functions of continuous spectrum to the
unit amplitude of incident wave. It is sufficient to calculate the
tunneling transparency for the barrier of specified shape. However,
to find the self-consistent solution taking into account the Coulomb
interaction of charge carriers one has to know the spatial
distribution of electrons, which is impossible without calculating
the continuous-spectrum eigenfunctions normalized to the
delta-function of quantum numbers.

Let us adduce the relations between the reflectivity $r$ and the
transparency $t$ of the barrier, which are necessary to carry out
the calculations:
\begin{equation}
\frac{1}{\partial q/\partial k}\left| t_{k}^{\rm R}\right|
^{2}=1-\left|  r_{k}^{\rm R}\right|^{2},\,\,\,
 \frac{1}{\partial k/\partial q}\left| t_{q}^{\rm L}\right|^{2}=
 1-\left| r_{q}^{\rm L}\right|^{2}.
\label{Eq t2-r2-R,L}%
\end{equation}
\begin{equation}
t_{q(k)}^{\rm L}=\frac{1}{\partial q/\partial k}t_{k}^{\rm R},
\,\,\,r_{q(k)}^{\rm L}=-r_{k}^{\rm R\ast}\left( t_{k}^{\rm R}
/t_{k}^{\rm R\ast}\right).\label{Eq (t,r)L-(t,r)R}%
\end{equation}
Eqs. \eqref{Eq (t,r)L-(t,r)R} are written for the case of left-hand
states normalized to $\delta(k-k')$, which is indicated by explicit
dependence $q(k)$ in the arguments.

The electron wave function in semiconductor can be represented as a
superposition of two real linearly independent solutions $\phi_1$
and $\phi_2$, which at $z \to \infty$ have the following asymptotic
behavior:
\begin{equation} \label{Eq phi asymptotic}
\begin{array}{l}
\phi_{k \tt 1}(z) = \sin (kz + \gamma_k),\\
\phi_{k \tt 2}(z) = \cos (kz + \gamma_k),\\
\end{array}
\end{equation}
where $\gamma_k$ is the phase of the wave function of k-th state.
With these functions, using boundary conditions and normalization,
one can build $\psi^{\rm R}_{k}(z)$, which, in its turn, is
sufficient to define $\psi^{\rm L}_{q}(z)$.

The solutions $\phi_1$ and $\phi_2$ are found by means of
integration of the Cauchy problem for the Schr\"{o}dinger equation
(\ref{schred^i}). With argument $z$ receding from the interface into
the bulk, functions $\phi_1$ increase exponentially in the
under-barrier region while $\phi_2$ decrease. If the potential
barrier is high enough, searching for the exponentially vanishing
solution presents some calculation difficulties. To overcome them,
the following technique was employed. Function $\phi_1$ is found
from the Cauchy problem solution with the initial conditions
$\phi_{k1}(0)= 1$ and $\phi'_{k1}(0)=\sqrt{U_{0}-k^2/2}$ specified
at the metal-semiconductor interface $z=0$. Eq. (\ref{schred^i}) is
integrated from $z=0$ to $z_\infty$ into the semiconductor bulk
where the effective potential becomes constant to a sufficient
degree of accuracy. The solution obtained is normalized to the unit
amplitude by formula similar to Eq. \eqref{Eq amplit-def}, and phase
$\gamma$ in expressions \eqref{Eq phi asymptotic} is found with the
help of relation \eqref{Eq phase-def}. It should be noted that
choice of initial conditions for $\phi_1$ is quite arbitrary,
because the normalization procedure provides the correct asymptotic
behavior of solutions at infinity. To find $\phi_2$, one also solves
the Cauchy problem, but the initial conditions are imposed this time
at the point $z=z_\infty$:
\begin{equation}\label{Eq phi2-Cauchy cond}
\begin{array}{l}
 \phi_{k2}(z_\infty) = \cos (kz_{\infty} + \gamma_k)\\
 \phi'_{k2}(z_\infty) = -k\phi_{k1}(z_\infty)\\
\end{array}
\end{equation}
and the integration is carried out from the semiconductor bulk to
metal boundary in the direction of the $\phi_{k2}(z)$ growth in the
barrier region thus improving the stability of procedure. The
accuracy criterion for the pair of solutions obtained is the
constancy of the Wronskian $W(\phi_{k1},\phi_{k2})$ at the interval
$[0,z_\infty]$.

Further on we assume that the effective masses of electrons in
semiconductor and metal are equal. Then both the wave function
itself and its first derivative are continuous at the boundary of
the two media. Of course, such premise is rather simulative, but the
complicated problem of deriving the matching conditions for the
envelope wave functions at the metal-semiconductor interface is not
the subject of our work. A different type of boundary condition at
the interface may, of course, affect the particular numerical
results, but this will hardly require a change in the iteration
scheme suggested for solving the Kohn-Sham equations for the
structures with the current states. The reasons for such opinion
will be discussed at the end of this section. Note also that
problems close to those considered here arise as well in the theory
of resonant tunneling diodes (see, for example,
\cite{Ferry-Good97}).

\subsection{Some calculation details}\label{Ssec_Schottky details}
Now then, on the basis of two functions \eqref{Eq phi asymptotic} we
should form the wave function $\psi^{\rm R}_{k}(z)$ of the
"right-hand" state that describes the tunneling of electrons from
the right-hand (semiconductor) electrode into the left-hand one,
which we still call for convenience "the metal". The asymptotic
boundary conditions \eqref{Eq psi_R-def} defining $\psi^{\rm
R}_{k}(z)$ realize the requirement to have a purely outgoing wave in
the metal and a linear combination of incident and reflected waves
in the semiconductor. These conditions can be used to find the
coefficients in the expansion
\begin{equation}\label{psiR as sum -def}
    \psi^{\rm R}_{k}(z)=\alpha_k \phi_{k1}(z)+\beta_k%
    \phi_{k2}(z).
\end{equation}
When $\psi^{\rm R}_{k}$ is found, $\psi^{\rm L}_{k}$ in general form
can be represented as a linear combination of two independent
solutions $\psi^{\rm R}_{k}$ and $\psi^{\rm R*}_{k}$. But to
calculate the electron density in the semiconductor, which is what
we only need for the problem to be solved, one has to know
$\psi^{\rm L}_{k}(z)$ only at $z \geq 0$. So to express $\psi^{\rm
L}_{k}$ through $\phi_{k1,2}$ it is enough to take advantage of the
asymptotic form \eqref{Eq phi asymptotic} of the left-hand state at
$z\rightarrow\infty$. All the above considering, one easily obtains
\begin{equation}\label{psiL-right represent}
    \psi^{\rm L}_{q(k)}(z)=t^{\rm L}_{q(k)}\exp(-{\rm i}\gamma_k)({\rm i}%
\phi_{k1}(z)+\phi_{k2}(z)).
\end{equation}

The problem of defining $\psi^{\rm R}_{k}(z)$ reduces to finding
four complex coefficients
$\alpha_k$,$\beta_k$, $r^{\rm R}_{k}$, $t^{\rm R}_k$. The four
corresponding equations are obtained by equating the asymptotic
formulae for function $\psi^{\rm R}_{k}$ and its derivative at
$z\rightarrow\infty$ to the asymptotic form of the linear
combination required, which provides two equations, and also from
two matching conditions for the function and its derivative at the
boundary $z=0$. In so doing, we assume that in the metal
\begin{equation}\label{psiR-left asym}
\psi^{\rm R}_{k}(k z)=t^{\rm R}_{k}\exp(-{\rm i} q(k) z).
\end{equation}
The calculation is carried out under the condition $U_{\rm M} < 0$,
which provides that the conduction band bottom in the left-hand
electrode is lower than that in the right-hand one, so that
inequality $q(\varepsilon,\mathbf{k}_{\|}) \neq
k(\varepsilon,\mathbf{k}_{\|})$ remains valid in spite of the
assumed equality of effective masses in the electrodes. At the same
time, the condition of effective mass equality ensures that function
$q(k)$ is independent of $\mathbf{k}_{\|}$, which reduces the
computational cost.

With some simple transformations, the sought system of linear
algebraic equations takes the form
\begin{align}
&\alpha\phi_{k1}(0)+\beta\phi_{k2}(0)-t_{k}=0, \nonumber\\
&\alpha\phi_{k1}^{\prime}(0)+\beta\phi_{k2}^{\prime}(0)+{\rm i}%
qt_{k}=0, \nonumber\\
&{\rm i}\alpha+\beta=2\exp({\rm i}\gamma_{k}), \nonumber\\
&{\rm i}\alpha-\beta+2r_{k}\exp(-{\rm i}\gamma_{k})=0. \label{Eq set}%
\end{align}
For brevity, we omit everywhere the index $\rm R$ of the right-hand
state, the wave-number index $k$ of coefficients $\alpha \, ,\beta$,
and imply that $q=q(k)$.

The solution of system \eqref{Eq set} can be represented as
\begin{align}
r_{k}  & =-\frac{\left(  q\Phi_k+{\rm i}\Phi_k^{\prime}\right)^{\ast}}%
{q\Phi_k+{\rm i}\Phi_k^{\prime}}, \label{solution-r}\\
t_{k}  & =\Phi_k^{\ast}+r_{k}\Phi_k\label{sol t-r_k},\\
\alpha_k & =-{\rm i}\left( \exp({\rm i}\gamma_k)-r_{k}\exp(-{\rm i}%
\gamma_k) \right),  \label{sol alpha-r_k}\\
\beta_k & =  \exp({\rm i}\gamma_k)+r_{k}\exp(-{\rm i}%
\gamma_k)  \label{sol beta-r_k},
\end{align}
where for compactness we introduced functions
\begin{align}
\Phi_k & =\exp(-{\rm i}\gamma_k)\left(  \phi_{k2}(0)+{\rm i}\phi
_{k1}(0)\right),  \label{Phi(0)-def}\\
\Phi_k^{\prime}  & =\exp(-{\rm i}\gamma_k)\left(
\phi_{k2}^{\prime}(0)+{\rm i}\phi_{k1}^{\prime}(0)\right)
\label{Phi'(0)-def},
\end{align}
and $\phi'_k$ means the derivative by $z$.

The scheme presented for calculating wave functions describes a
modified procedure for solving the Schr\"{o}dinger equation
\eqref{Schred^i-cycle} for unbounded system as distinct from a
semi-bounded one (see Section \ref{SSec-Algorithm/Applicat}). The
corresponding formula to replace Eq. \eqref{n^i-calc} for
calculating the electron density in a barrier structure is found
from the general expression derived in \cite{ASH06}. Taking into
account that the electron density should be known only at $z \geq
0$, and the absence of bias voltage, we get
\begin{equation}
N(z)=2\int_{0}^{\infty}dk\int_{-\infty}^{\infty}d\mathbf{k}_{\|}%
f\left(\varepsilon(k,\mathbf{k}_{\|})\right) \left[ \left
|\Psi_{q(k)}^{\rm L} (z) \right|^{2} + \left |\Psi_{k}^{\rm R}(z)
\right|^{2}\right],
\label{Eq N(z)-Psi_Schot}%
\end{equation}
where $f(\varepsilon)$ is the Fermi distribution function, and
energy eigenvalues $\varepsilon(k,\mathbf{k}_{\|})$ are given by
formula \eqref{Eq k-q def}. The Fermi level position is defined by
the neutrality condition in the semiconductor bulk at $z \to
\infty$. Using the definitions \eqref{Eq Psi_R-def}-\eqref{Eq
Psi_L-def}, \eqref{Eq C4L-k norm}, and \eqref{psiR as sum
-def}-\eqref{psiL-right represent} along with relations \eqref{Eq
t2-r2-R,L}-\eqref{Eq (t,r)L-(t,r)R} between the amplitude
coefficients of barrier reflection and transparency, we obtain from
\eqref{Eq N(z)-Psi_Schot} an expression for electron density in
terms of the two real solutions $\phi_{k1,2}(z)$ found numerically,
\begin{multline}\label{Eq N(z)-fin_Sch},
N(z)=\frac{2}{(2\pi)^3}\int_{0}^{\infty}dk\int_{-\infty}^{\infty}d\mathbf{k}_{\|}%
f\left(\varepsilon(k,\mathbf{k}_{\|})\right)\times\\
\left[ \left |\alpha_k \phi_{k1}(z) +\beta_k \phi_{k2}(z)
\right|^{2} + \left(1-\left|r_{k}^{\rm R} \right|^2\right) \left|
\phi_{k1}(z)-{\rm i}\phi_{k2}(z)\right|^2 \right].
\end{multline}
Here the coefficients $r_{k}^{\rm R},\,\alpha_k$, and $\beta_k$ are
defined in \eqref{solution-r}-\eqref{sol beta-r_k}.

The expression \eqref{Eq N(z)-fin_Sch} for electron density provides
an explicit representation of the formula \eqref{Ni_gen def} of
iteration procedure for the case under consideration.

\subsection{Results and discussion}\label{Sch_results}
Calculations were carried out at the values of parameters typical
for the Al/n-GaAs junctions ($m^{*}=0.07m_{\rm e}$, $\Phi_{\rm
s}=0.9$ eV, $\kappa=12.5$) with ionized donor concentration $N_{D} =
10^{18} {\rm cm}^{-3}$, which corresponds to $R_{\rm s}=0.66$. For
$q$, in virtue of the large difference between Fermi energies in
semiconductor and aluminium, it turned out possible to accept a
constant value that was chosen equal to the radius of the Fermi
surface of electron gas at $R_{\rm s}=2.07$, $\kappa_{\rm m}=1$.
\begin{figure}[h]\center
\includegraphics[scale=0.6,]{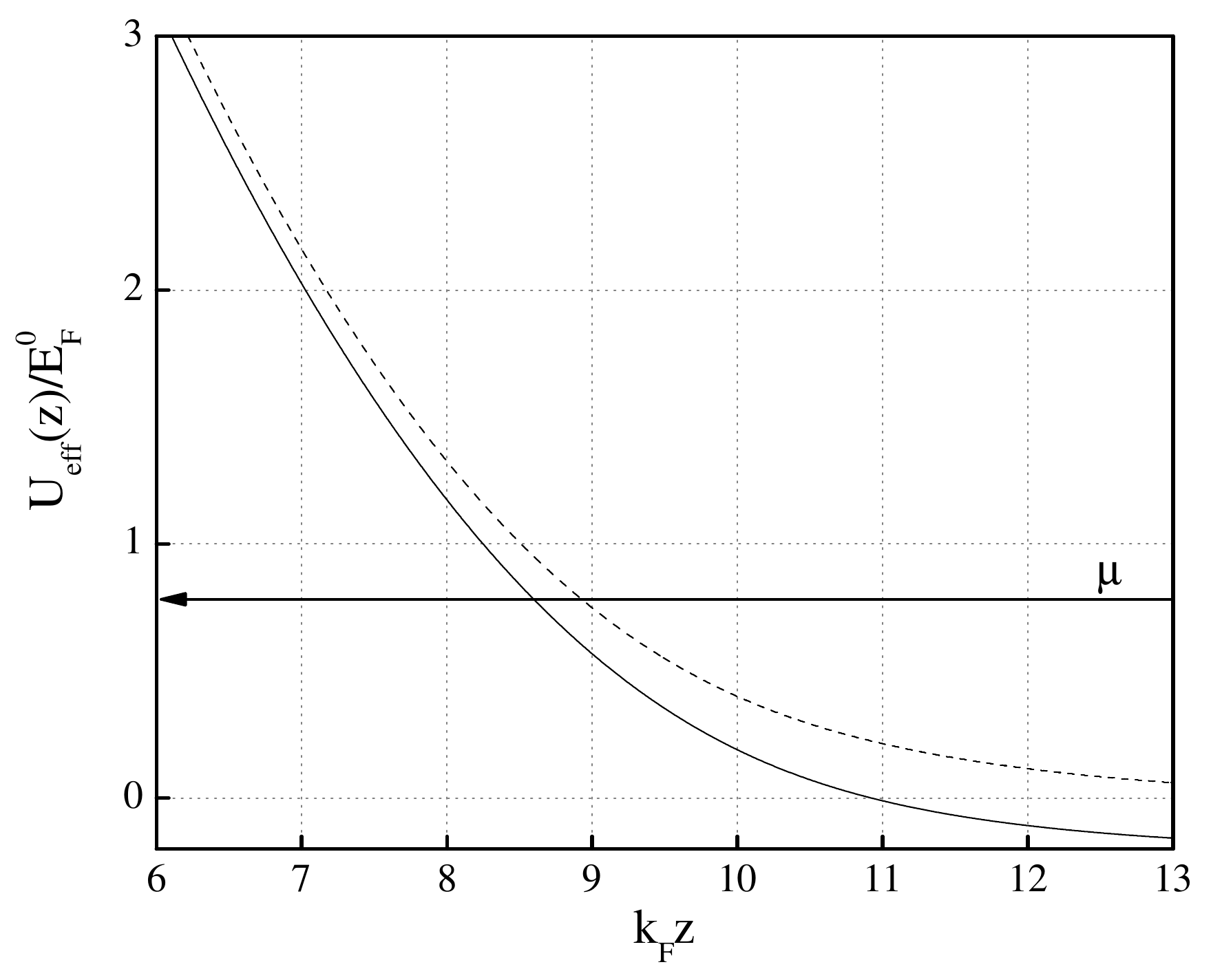}\\
  \caption{Self-consistent potential for a Schottky barrier structure ($R_s = 0.66$).
Solid line represents the potential found with account of exchange-correlation interaction,
dashed line is the potential in Hartree approximation.
The Fermi level position with account of $U_{\rm xc}$} is marked.
 \label{fig8-Schott barr}
\end{figure}
Fig. \ref {fig8-Schott barr} represents the coordinate dependence of
the self-consistent effective potential calculated with account of
the exchange-correlation potential of the electrons (curve 1) and in
Hartree approximation (curve 2) near the metal-semiconductor
interface. One can see that including $U_{\rm xc}$ into calculation
modifies the potential barrier shape making the drop of potential
steeper. This difference has a notable impact on the shape of
theoretical current-voltage curve of tunneling junction,
particularly in the case of forward bias when electrons tunnel from
semiconductor into metal through the effective potential barrier
region below the Fermi level (see \cite{ASH-Zai_SSC76},
\cite{Kot-Ash_88}-\cite{Ash-Kot_01JETPL}). It is demonstrated in the
same works that the account of $U_{\rm xc}$ allows describing more
accurately the current-voltage curves of real structures. Hence, the
representation of the exchange-correlation interaction of electrons
by means of the local density approximation remains quite accurate
for the electron states located sufficiently deep under the Fermi
level.

Besides, the use of self-consistent solution for potential in the
tunneling current formula makes possible to define the parameters
$\Phi_{\rm s}$ and $N_D$ of the metal/heavily n-doped GaAs junction
from the bias voltage dependence of differential resistance
\cite{Kot-Bej-Ash_85FTT}-\cite{Dizh-ea_2001}\footnote{The Fig.3
missed in the published version of the Ref. \cite{Dizh-ea_2001} is
restored in the arXiv preprint.}. The Schottky barrier was
calculated there in the Thomas-Fermi and Thomas-Fermi-Gombas
approximations. Though in such structures the conditions are well
fulfilled for the Schottky barrier to be quasiclassical, i. e.
$k_{\rm F}L \gg 1$ and $k_{\rm F}l_{\rm TF} \gg 1$, where $L$ and
$l_{\rm TF}$ denote the characteristic scales of spatial variations
of the self-consistent potential inside and outside the barrier
\cite{Kot-Bej-Ash_85FTT}, the question of a possible notable change
in barrier shape beyond the scope of the Thomas-Fermi approximation
still remained. The exact calculation showed that the large-scale
behavior of the Schottky-barrier effective potential is accurately
described by the potential found in the Thomas-Fermi approximation,
and the agreement between them improves with decreasing $R_{\rm s}$.

\begin{figure}[htb]\center
    \includegraphics[scale=0.6,]{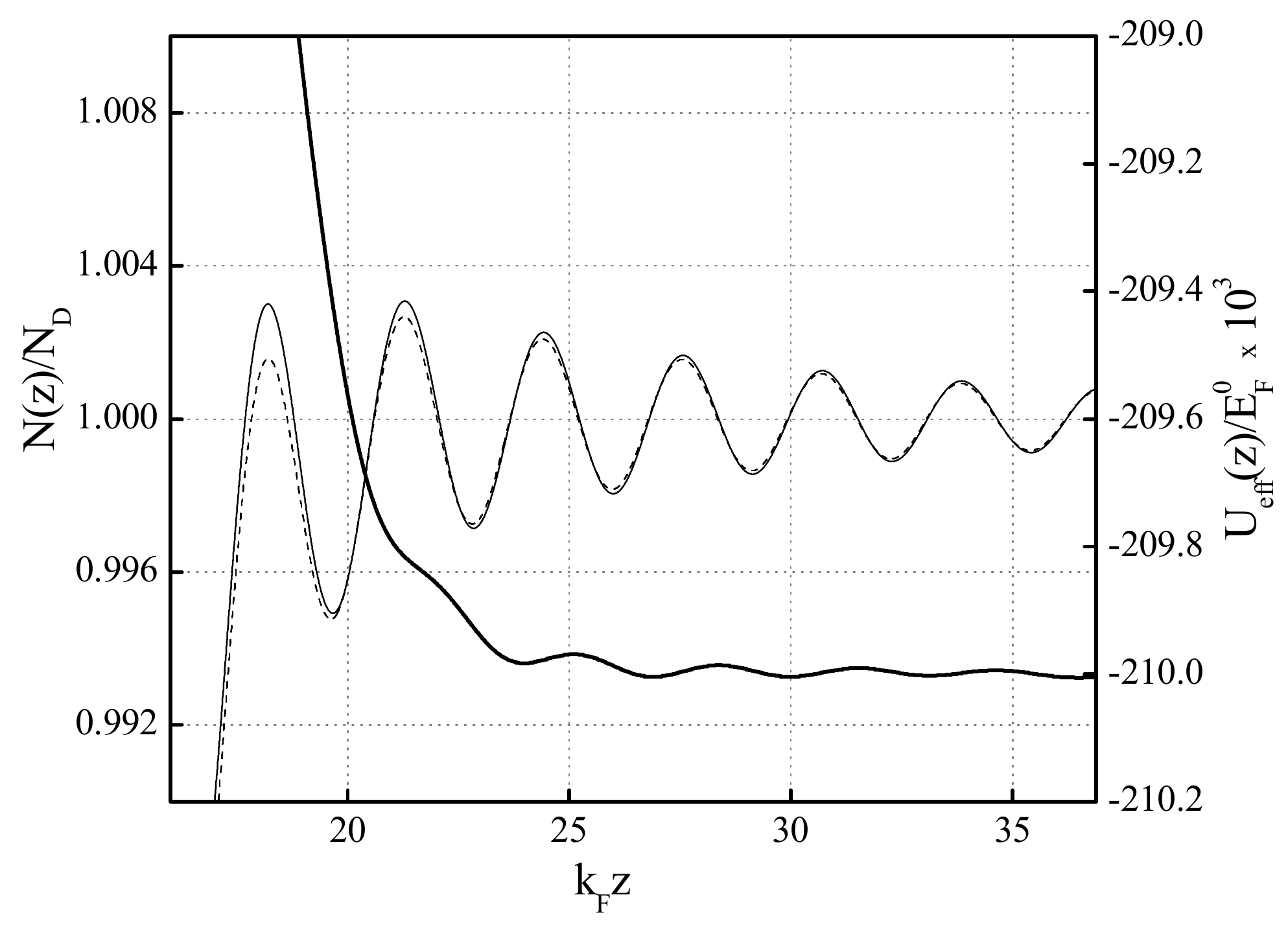}\\
  \caption{Friedel oscillations of electron density and
  effective potential in the Schottky barrier structure
  at $R_{\rm s}=0.66$. Solid line is electron density with account of
  exchange-correlation interaction, dashed line is
  electron density in the Hartree approximation.
  Bold solid line is the self-consistent effective potential.}
  \label{fig9_Schott-osc}
\end{figure}
In the semiconductor bulk, the coordinate dependence of
self-consistent potential and electron density manifests
oscillations (Fig. \ref {fig9_Schott-osc}), which are absent in the
Thomas-Fermi approximation. As a whole, the oscillations caused by
Schottky barrier are less pronounced than those in the case of an
infinitely high potential wall (compare with Fig.
\ref{fig4_friedel}), which is apparently related to the
quasiclassical smoothness of the Schottky barrier. This is
particularly true in regard to the effective potential oscillations.
As can be seen in Fig. \ref {fig9_Schott-osc}, the oscillation
magnitude of effective potential is quite small, of the order of
$10^{-4}$ and less, while the electron density oscillations in the
exact solution are, on the contrary, more intensive than those in
the Hartree approximation. The suppression of the effective
potential oscillations compared to those of its Coulomb component
was already discussed in Section \ref{Subsec_self-levels}. The
amplification of density oscillations in the exact solution with
respect to solution in the Hartree approximation can be explained by
analyzing the linearized expression for the induced density.

It is well known that linearization of formula \eqref{N_ind-def} in
the Hartree approximation at small values of potential $U$ leads in
the self-consistent Poisson equation \eqref{Poiss-N_ind} to the
expression for the density (charge) increment, which is generally
written as $4 \pi \delta N_{\rm ind}=-k_{\rm TF}^2 U$. The account
of the electron density dependence of exchange-correlation potential
leads to a similar relation $4 \pi \delta N_{\rm ind}=-k_{\rm scr}^2
U$,
the inverse square screening length being
\begin{equation}\label{Eq screen-length_def}
    k_{\rm scr}^2=\frac{k_{\rm TF}^2}{1+\frac{3}{2}%
    du_{\rm xc}(n)/dn}.
\end{equation}
The dimensionless exchange-correlation potential grows in absolute
magnitude with decreasing electron concentration, so the derivative
$du_{\rm xc}/dn < 0$, and the denominator in formula \eqref{Eq
screen-length_def} tends to zero with growing $R_{\rm s}$. The
equality of the denominator to zero provides the critical value of
$R_{\rm sc}$. So, at comparable values of the self-consistent
potential oscillations in the exact solution and that obtained in
the Hartree approximation, which is in evidence, the electron
density oscillations in the exact solution will be amplified
increasingly with $R_{\rm s}$ getting closer to $R_{\rm sc}$.
Accordingly, the screening region, where the perturbation drops
exponentially, will be reduced.

Summarizing the results of analyzing the Friedel-type oscillations
caused by the violation of homogeneity of degenerate electron gas,
we note that the account of exchange-correlation interaction of
electrons leads to the increase of electron density oscillations and
the decrease of effective-potential oscillations with respect to
results obtained in the Hartree approximation.

To conclude this section, let us discuss to what extent the results
obtained depend on the adopted condition of equal effective masses
in the two electrodes. Generally speaking, the problem of boundary
conditions for the envelope function at the interface of two solids
differing not only by effective masses, but also by the type of
charge carriers and their dispersion law, is rather complicated.
According to the current opinion, such conditions cannot be reduced
to a universal relation between the values of wave function and its
derivative on both sides of the interface, but should be derived for
each specific case (see, for example, the critical discussion
related to heterojunctions in \cite{Takh-Volk_99JETP},
\cite{Tokatly-ea_02PRB65}).

Staying within the framework of phenomenological description of
boundary conditions, one can write them in the most general form in
terms of the transfer matrix $\mathbb{T}$, the elements of which are
constrained by the requirement of Hermiticity of the Hamiltonian
\cite{Tokatly-ea_02PRB65}. If envelope-function equations for both
electrodes can be taken in the single-band approximation, the
boundary conditions are described by a $2\times2$ matrix. In the
case when the $\mathbb{T}$-matrix can be diagonal
\cite{Ando-ea_89PRB40}, the final formulae
\eqref{solution-r}-\eqref{sol beta-r_k} keep their form with the
barrier transparency $t_k$ and wave vector $q$ replaced by
renormalized values $\tilde t=T_{11}t_k$ and $\tilde%
q=(T_{22}/T_{11})q$.

If matrix $\mathbb{T}$ has the nonzero off-diagonal elements, it
does not lead to crucial difficulties in building the R- and
L-solutions of the Schr\"{o}dinger equation according to their
asymptotic behavior at $z \to \pm \infty$ from the particular real
solutions. However, the analysis of the most general case in this
work would make the statement of fundamental aspects of iteration
algorithm application unreasonably cumbersome. At the same time, the
inequality of wave vectors $q(\varepsilon)\neq k(\varepsilon)$
typical for the non-symmetrical barrier structures was necessary to
include into the calculation.
\section{Conclusion}
The realized iteration algorithm for solving the self-consistent
field equations was found to converge in all the cases considered of
typical inhomogeneous electron systems with continuous spectrum. The
difficulties of its application arose only when considering the
jellium model at large values of $R_{\rm s}\geq 5$ that do not occur
in most metals and semiconductor structures. It was also shown that
an explicit account of screening properties of electron gas in the
algorithm for numerical analysis of infinite Coulomb systems allows
to consider a selected limited volume of such system, transferring
the boundary conditions from infinity to the surface and assuming
the energy spectrum of single-electron states to be strictly
continuous. Such approach also eliminates the difficulties that
arise when considering an inhomogeneous infinite system with
continuous spectrum as a limit of a finite one with quasi-discrete
spectrum.

This work was partially supported by Russian Foundation for Basic
Research with grants 07-02-01481 and 06-02-16955.

\end{document}